\documentclass[journal,10pt]{IEEEtran}

\usepackage{cite}
\usepackage{color} 
\usepackage{bm}
\usepackage{subfigure}
\usepackage{multirow}
\usepackage{tabularx}

\ifCLASSINFOpdf
   \usepackage[pdftex]{graphicx}
\else
\fi
\usepackage{cite}
\usepackage{amsmath,amssymb,graphicx,algorithm,algorithmic}
\usepackage{makecell}

\usepackage{color}   
\usepackage{array}
\begin{document}

\title{Channel Measurements and Modeling for Dynamic Vehicular ISAC Scenarios at 28 GHz}

\author{Zhengyu~Zhang,~\IEEEmembership{Student Member, IEEE,}
        Ruisi~He,~\IEEEmembership{Senior Member, IEEE,}
        Bo~Ai,~\IEEEmembership{Fellow, IEEE,}
        Mi~Yang,~\IEEEmembership{Member, IEEE,}
        Xuejian Zhang,
        Ziyi Qi
        and Yuan Yuan
\thanks{


Zhengyu Zhang, Ruisi He, Bo Ai, Mi Yang, Xuejian Zhang, Ziyi Qi and Yuan Yuan are with the School of Electronics and Information Engineering, Beijing Jiaotong University, Beijing 100044, China.

              }
\thanks{
              }}

\markboth{IEEE xxx,~Vol.~XX, No.~XX, XXX~202x}
{}

\maketitle

\begin{abstract}

Integrated sensing and communication (ISAC) is a promising technology for 6G, with the goal of providing end-to-end information processing and inherent perception capabilities for future communication systems. Within ISAC emerging application scenarios, vehicular ISAC technologies have the potential to enhance traffic efficiency and safety through integration of communication and synchronized perception abilities. To establish a foundational theoretical support for vehicular ISAC system design and standardization, it is necessary to conduct channel measurements, and modeling to obtain a deep understanding of the radio propagation. In this paper, a dynamic statistical channel model is proposed for vehicular ISAC scenarios, incorporating Sensing Multipath Components (S-MPCs) and Clutter Multipath Components (C-MPCs), which are identified by the proposed tracking algorithm. Based on actual vehicular ISAC channel measurements at 28 GHz, time-varying sensing characteristics in front, left, and right directions are investigated. To model the dynamic evolution process of channel, number of new S-MPCs, lifetimes, initial power and delay positions, dynamic variations within their lifetimes, clustering, power decay, and fading of C-MPCs are statistically characterized. Finally, the paper provides implementation of dynamic vehicular ISAC model and validates it by comparing key simulation statistics between measurements and simulations.

\end{abstract}


\IEEEpeerreviewmaketitle

\section{Introduction}

\IEEEPARstart{W}{i}th the development of wireless communication systems, the sixth generation (6G) network not only represents an enhancement of current communication technologies, but also possesses the capability to provide ubiquitous sensing, seamless connectivity and advanced intelligence through wireless devices\cite{1}. In this vision, higher demands for end-to-end information processing ability and native perception ability of 6G networks are required, so as to prompt the recent research interest in integrated sensing and communication (ISAC) techniques\cite{2}. Technically, ISAC can enhance spectrum and energy efficiency while reducing hardware and signal processing costs, which is achieved by enabling the integration of sensing and communication functionalities within a single transmission, device, or network infrastructure\cite{3}. ISAC has been expected to enable communication-assisted sensing or sensing-assisted communication, which enhances jointly sensing and communication applications in 6G scenarios, including smart home systems, enhanced positioning and tracking, human-computer interaction, environmental monitoring, sensing-assisted beam training, and so on\cite{4}. 

Recently, ISAC techniques have emerged as a pivotal component in the vehicle industry, playing a crucial role in enhancing vehicle safety and performance\cite{5,CJE1,7}. Under the background of deep integration of communication and sensing technologies in the vehicle industry, vehicular ISAC networks is one of the most potential ISAC application scenarios. On the one hand, lots of communication transceivers and sensors are equipped on the autonomous vehicles \cite{8}. For example, the novel vehicles is usually equipped with eight cameras for a 360-degree of environment perception\cite{9}. On the other hand, for vehicular ISAC, it is possible to reuse the current dense deployment road-side units (RSUs) for sensing with only minor modifications in hardware, signaling strategy, and communication standards\cite{CJE2,11,12}. In this way, vehicular ISAC can connect vehicles with surrounding vehicles, traffic infrastructures, and networks, as well as use the integration gain of ISAC to obtain more accurate perception information and better wireless communication quality, so as to reduce traffic jams and accident rates efficiently.

However, vehicular ISAC technologies are in its early stages, and a deep understanding of the radio propagation mechanism of wireless channel is crucial for its design, performance evaluation, and standardization\cite{add3,14,15}. In contrast to traditional communication channels, vehicular ISAC channels integrate both sensing and communication channels. This integration encompasses radio propagation from the communication transmitter to the communication receiver, as well as echo propagation from the sensing transceiver to scatterers and back to the sensing transceiver\cite{16,17}. Therefore, vehicular ISAC channels are more sensitive to the driving environment, such as surrounding vehicles, people and other scatterers. This leads to the presence of abundant effective sensing multipaths as well as clutters, with a more rapid birth and death process for them. Moreover, owing to the dynamics of these scatterers and the mobility of vehicles, time-varying behavior is a crucial characteristic of vehicular ISAC channels, which should be sufficiently incorporated into channel modeling. Besides, during the driving process of vehicles, the road conditions and environmental distributions vary across different sensing directions, such as the types of scatterers, time-varying behavior, etc., resulting in distinct channel characteristics, it is necessary for different directions to model vehicular ISAC channels accordingly. 

Therefore, dynamic vehicular ISAC channel modeling is a challenging task, the professional channel measurements and characterization are necessary for accurate modeling\cite{add4}. Over the past few years, extensive research has been conducted on vehicular channel measurements and models, with a predominant focus on communication channels where transmitters and receivers are mounted in different positions, and these studies are mostly conducted in sub-6 GHz frequencies rather than millimeter waves, which are more attractive for ISAC technologies. For example, Ref.\cite{18} presented a MIMO model for Vehicle-to-Vehicle (V2V) channels based on measurements performed at 5.2 GHz in highway and rural environments. An extensive communication channel measurements are conducted in a straight subway tunnel at 1.8 and 5.8 GHz in Ref.\cite{19} and on a suspension bridge and a beam bridge at 5.9 GHz in Ref.\cite{20}, respectively. Measurement results and propagation channel model are presented in Ref.\cite{add2}, in which a bus acts either as a shadowing object or as a relay between two passenger cars. Dynamic time-varying has been main characteristics in vehicular channels. Ref.\cite{21} carried out measurement campaigns and analyzed the non-stationarity of V2V channels as well as vehicle-to-infrastructure (V2I) channels. Ref.\cite{22} proposed a dynamic wideband directional channel model based on measurements conducted at 5.3 GHz in suburban, urban, and underground parking environments. By incorporating dynamic behaviors of the MPCs, the model is capable of handling time-varying characteristics. In Ref.\cite{23}, channel measurements at 5.9 GHz in street intersection scenarios are carried out and provide data for the characterization and modeling of time-varying vehicular channels. Ref.\cite{add1} proposed a geometric MIMO channel model, which is adaptable to a variety of millimeter-wave mobile-to-mobile scenarios. Although the aforementioned vehicular channel measurements covers most dynamic scenarios, these measurements and models cannot be generalized to vehicular ISAC channels. 

On the other hand, the literature on ISAC channel measurements and models is relatively scarce and mainly focuses on non-dynamic scenarios, resulting in a lack of time-varying characteristics. Ref.\cite{24} analyze the ISAC channel characteristics based on the measurements at 28 GHz in an outdoor scenario, where a metal board is placed in a stationary position as sensing target. Ref.\cite{25} compared ISAC channel characteristics between communication channels and sensing channels through actual measurements in outdoor static scenarios, and the correlation model of multipaths in ISAC channels are measured and characterized in Ref.\cite{26}. Ref.\cite{27} measured the azimuth power spectrum of vehicular ISAC channels, while the vehicles are moving slowly due to narrow experimental space. Ref.\cite{28} conducted communication and sensing channel measurements for an indoor static scenario at 28 GHz. These static measurement scenarios cannot reflect the time-varying characteristics of ISAC channels. Besides, ISAC channel models for various application scenarios have also been widely researched. In Ref.\cite{29}, a vehicle ISAC system based on IEEE 802.11ad is presented, where communication channel is considered as a Rice channel and sensing channel is considered as a path-based model. Ref.\cite{30} proposed a scatterer-based hybrid channel model framework, which first generate scatterers in the propagation environment and then use simplified ray-tracing to generate channels. In Ref.\cite{31}, a novel ISAC channel model combining forward and backward scattering is proposed, where the correlations are investigated. In Ref.\cite{32}, a unified approach to model an ISAC channel is proposed, where the pathloss, LOS probability, Doppler and the fast fading are provided systematically. However, similar to ISAC channel measurements, these models is difficult to represent dynamic evolution over time, such as the lifetimes of multipaths, making them unsuitable for application in vehicular ISAC scenarios. To the best of the authors' knowledge, there exist few actual channel measurements and modeling for dynamic vehicular ISAC scenarios.

As mentioned earlier, despite numerous efforts in measuring and modeling of vehicular channels, these studies have primarily focused on communication channels, lacking a comprehensive representation of the characteristics of sensing channels, especially their dynamic characteristics. To fill the gap, this paper conducts vehicular channel measurements from the perspective of sensing and proposes a measurement-based dynamic statistical channel model aiming for vehicular ISAC scenarios. The main contributions are summarized as

\begin{itemize}
\item[1)]
A novel dynamic ISAC channel tapped-delay line model is proposed. In the proposed model, multipaths are categorized into two groups: sensing multipath components (S-MPCs) and clutter multipath components (C-MPCs), each described with its own statistical distribution.
\end{itemize}

\begin{itemize}
\item[2)]
A vehicular ISAC channel measurement system is used to conduct dynamic channel measurements at 28 GHz in different directions, including front, left and right directions. The transmitter and receiver both mounted on a vehicle to focus on the time-varying sensing characteristics.

\end{itemize}

\begin{itemize}
\item[3)]
A tracking algorithm is utilized to identify S-MPCs and C-MPCs, and a series of characterization, including number of new S-MPCs, lifetimes, initial power and delay positions, dynamic variations within their lifetimes, clustering, power decay and fading of C-MPCs, is statistically modeled to enable simulated channel with dynamic evolution process.

\end{itemize}

\begin{itemize}
\item[4)]
Comparisons of channel characteristic parameters between measurements and simulations are conducted to implement and validate the model, which reflect the accuracy and effectiveness of the proposed model.

\end{itemize}

The remainder of this paper is structured as follows. In section II, a dynamic channel model is proposed for vehicular ISAC scenarios. Section III introduces the vehicular ISAC channel measurement system and measurement campaign. In section IV, the data processing for modeling are presented, including S-MPCs and C-MPCs modeling. Section V presents the channel model implementation and validation. Finally, the conclusion is given in Section VI.

\section{Dynamic ISAC Channel model}

To model vehicular ISAC channels with time-varying characteristics, we introduce a novel dynamic ISAC channel tapped-delay line model in this paper, which is composed of S-MPCs and C-MPCs. For S-MPCs, they typically exhibit higher power and more pronounced clustering, attributed to the strong reflection from sensing targets, and occur continuouly over time. In contrast, C-MPCs have lower power and are generally distributed across the entire delay domain due to widespread scatterers and noise, and occur randomly over time. The ISAC channel impulse response can be expressed as

\begin{equation}
\begin{split}
&h(t, \tau)= \\
&\sum_{l=1}^{L(t)} a_l(t) e^{j \varphi_l} \cdot \delta\left(\tau-\tau_l(t)\right)+\sum_{n=1}^{N(t)} a_n(t) e^{j \varphi_n} \cdot \delta\left(\tau-\tau_n(t)\right) 
\end{split}
\end{equation}
where $\delta(\cdot)$ is the Dirac delta function, $L(t)$ is the number of effective S-MPCs, $N(t)$ is the number of C-MPCs, $a_l(t)$, $\tau_l(t)$ as well as $a_n(t)$, $\tau_n(t)$ are the amplitude and delay of the $l$th and $n$th MPCs at time $t$ respectively. Noted that $a_l(t)$, $\tau_l(t)$ and $a_n(t)$, $\tau_n(t)$ are time-varying parameters. $\varphi_l$ and $\varphi_n$ are the phase and are assumed to be uniformly distributed in the range of $[0,2\pi]$.

For time-varying vehicular ISAC channels, the behavior of S-MPCs changes continuously over time, while C-MPCs remain time-varying and independent at each time. Specifically, due to vehicular movement and dynamic scatterers, each S-MPC has own lifetime that appear and disappear at random time instant, and the amplitude and delay are time-varying within lifetimes. Due to random and unknown environment, C-MPCs are modeled as a type of random distribution characterized by specific statistical properties. The set of all S-MPCs that exist at time instant $t_i$ are $L_i$, which consist of two sets of S-MPCs, as follows:
\begin{equation}
L_i=L_{1, i} \cup L_{2, i} \cup, \ldots, L_{j, i} \cup, \ldots, L_{i-1, i} \cup L_{i, i}
\end{equation}
where $L_{i,i}$ is the set of S-MPCs that first appear at time $t_i$, and $L_{j,i}$ is the set of S-MPCs that first appear at time $t_j$ and still exist at time $t_i$. The indexes of path in $L_{i,i}$ and $L_{j,i}$ are $l_{i,i}=1,2,...,N(t_{i,i})$ and $l_{j,i}=1,2,...,N(t_{j,i})$ respectively. And the number of all effective S-MPCs at time $t_i$ is expressed as
\begin{equation}
L\left(t_i\right)=L\left(t_{i, i}\right)+\sum_{j=1}^{i-1} L\left(t_{j, i}\right)(0<j<i)
\end{equation}

The channel impulse response at time $t_i$ can be reconstructed based on the collection of S-MPCs in all sets of $L_{i,i}$ and $L_{j,i}$, as well as C-MPCs across all delay bins. Considering the time-varying characteristics, channel impulse response can be thus expressed as
\begin{equation}
\begin{aligned}  
h\left(t_i, \tau\left(t_i\right)\right)= & \sum_{j=1}^{i-1} \sum_{l_{j, i}=1}^{L\left(t_{j, i}\right)} a_{l_{j, i}}\left(t_i\right) e^{j \varphi_{l_{j, i}}} \cdot \delta\left(\tau-\tau_{l_{j, i}}\left(t_i\right)\right)\\&+\sum_{l_{i, i}=1}^{L\left(t_{i, i}\right)} a_{l_{i, i}}\left(t_i\right) e^{j \varphi_{l_{i, i}} \cdot \delta\left(\tau-\tau_{l_{i, i}}\left(t_i\right)\right)} \\
& +a\left(t_i, \tau\left(t_i\right)\right) e^{j \varphi}
\end{aligned}
\end{equation}

The first term of equation (4) represents the old S-MPCs that appeared before time $t_i$, the second term of equation (4) represents the new S-MPCs that appeared at time $t_i$, and the third term of equation (4) represents the immediate C-MPCs that existed at time $t_i$ and delay $\tau(t_i)$. By modeling the “birth” and “death” processes of S-MPCs and the clutter floor of C-MPCs, the channel model can generate the dynamic and successive evolutions of impulse response at any time. In order to provide the proposed model, we statistically extract the following parameters from measurements:

$\bullet$ the length of S-MPCs lifetime;

$\bullet$ the number of newly appearing S-MPCs in each snapshot;

$\bullet$ the initial power and delay of newly appearing S-MPCs in each snapshot;

$\bullet$ the dynamic evolution of power and delay of S-MPCs over successive snapshots within the lifetime;

$\bullet$ the clustering of S-MPCs in each snapshot;

$\bullet$ the power decay and small-scale fading of C-MPCs in each snapshot.

Utilizing the statistical characterization of the aforementioned parameters, a dynamic ISAC channel model can be developed to simulate time-varying channels.

\section{Measurement Campaign}
\subsection{Measurement System}

\begin{figure*}[tbp]
\centering
\subfigure[]{\includegraphics[width=7in]{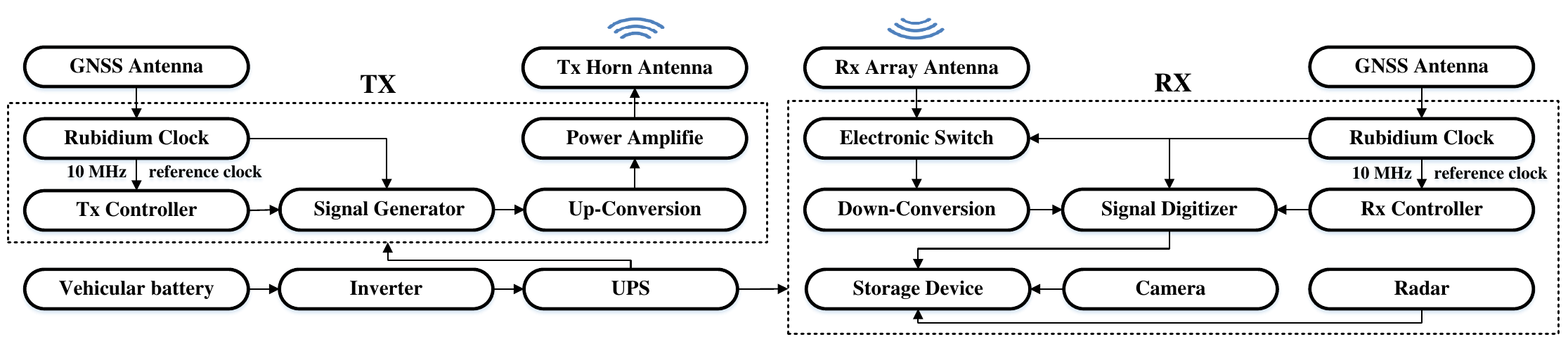}}

\caption{Vehicular ISAC channel measurement system.}
\label{fig}
\end{figure*}

\begin{figure*}[tbp]
\centering
\subfigure[]{\includegraphics[width=2.25in]{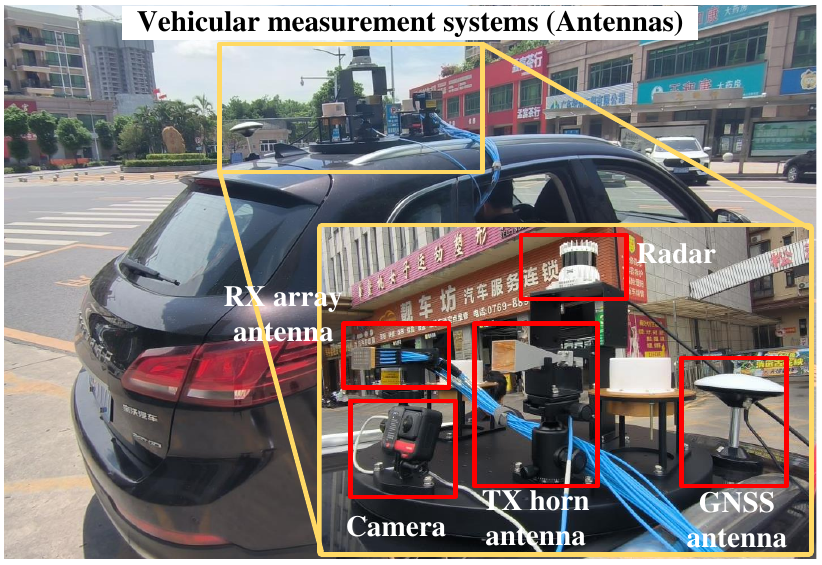}}
\subfigure[]{\includegraphics[width=2.25in]{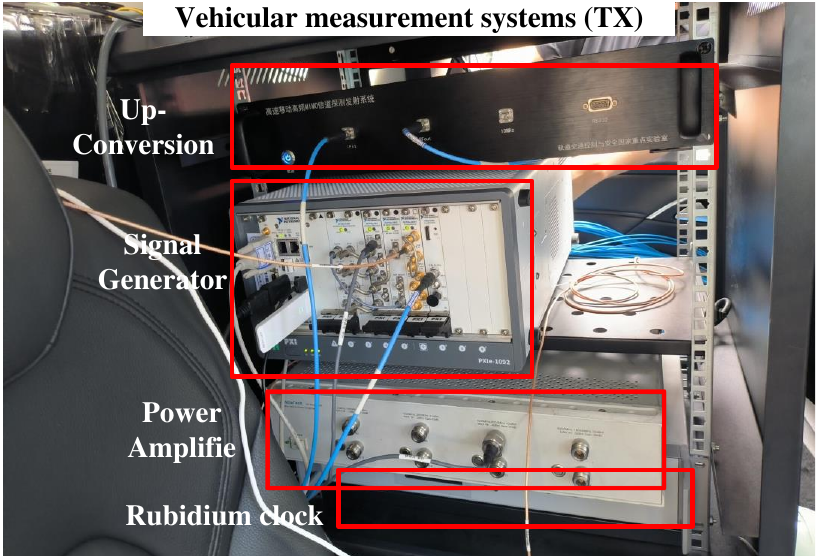}}
\subfigure[]{\includegraphics[width=2.2in]{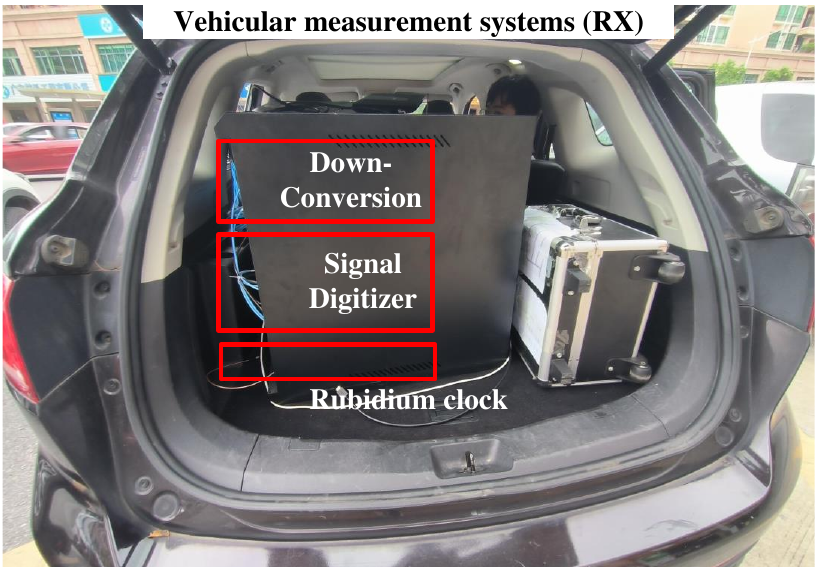}}
\caption{Key equipment of vehicular ISAC channel measurement system. (a) Antennas. (b) TX. (c) RX.}
\label{fig}
\end{figure*}

The vehicular ISAC channel measurement system is designed as shown in Fig. 1, including a signal generator based transmitter (TX), a signal digitizer based receiver (RX), and a power supply based on the vehicular battery. The architecture and key equipment of measurement system are illustrated in Fig. 1 and Fig. 2 respectively. For TX, the National Instruments (NI) PXIe-5745 is employed as ``signal generator", which generates the baseband signals with 1 GHz bandwidth. Then, baseband signals are converted to 28 GHz frequency through the up-conversion module, and transmitted through a power amplifier with gain of 28 dB and horn antennas with a directional beamwidth of 15 degrees and gain of 20 dB. For RX, sounding signals are received through array antennas and electronic switches. The array antenna is a 32-element rectangular array with gain of 5 dB, and different channels in the switches are utilized to enhance received signals. Then, they are converted to baseband through the down-conversion module and stored. During the measurement, RGB video data from the camera and point-cloud data from the radar are collected and stored simultaneously to aid in mapping between channel MPCs and environmental scatterers. The TX and RX are synchronized by the rubidium clocks amd Global Navigation Satellite System (GNSS) antennas, which can provide 10 MHz reference clock as well as vehicle's position. The inverter is employed to convert the vehicular 12V DC battery to 220V AC, and uninterruptible power supplies (UPS) is employed to provide stable power to the measurement equipment. The detailed configurations of vehicular ISAC channel measurement system are presented in Table I. Prior to measurements, back-to-back measurements are conducted to eliminate the impact of measurement equipment, such as cables, switches, transceivers, etc, thereby ensuring accurate measurement data.

\begin{table}[]
\centering
\caption{Configurations of ISAC measurement system.} \label{Table1Label}
\scalebox{1.2}{
\begin{tabular}{cc}
\hline
Parameters                 & Value                \\ \hline
Center frequency           & 28 GHz               \\
Bandwidth                  & 1 GHz                \\
Transmit power             & 28 dBm               \\
Delay resolution           & 1 ns                 \\
Sounding signal            & Multi-carrier signal \\
Transmitter antenna        & Horn antenna         \\
Receiver antenna           & 4$\times$8 array antenna\\    
Vehicle speed              & 30 km/h\\  
Transceiver height          & 2.3m \\ \hline
\end{tabular}}
\end{table}

\subsection{Measurement Scenarios}

\begin{figure}[tbp]
\centering
\subfigure[]{\includegraphics[width=3.5in]{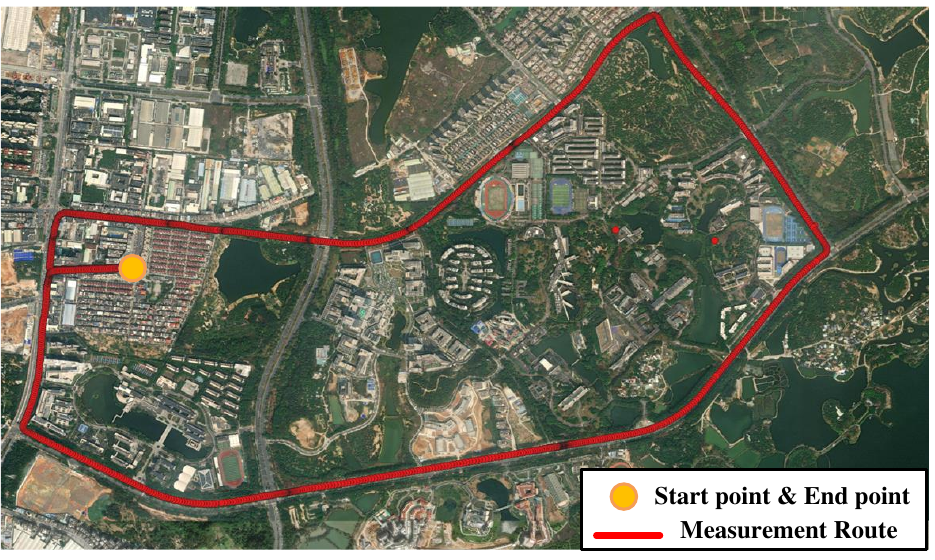}}
\subfigure[]{\includegraphics[width=3.5in]{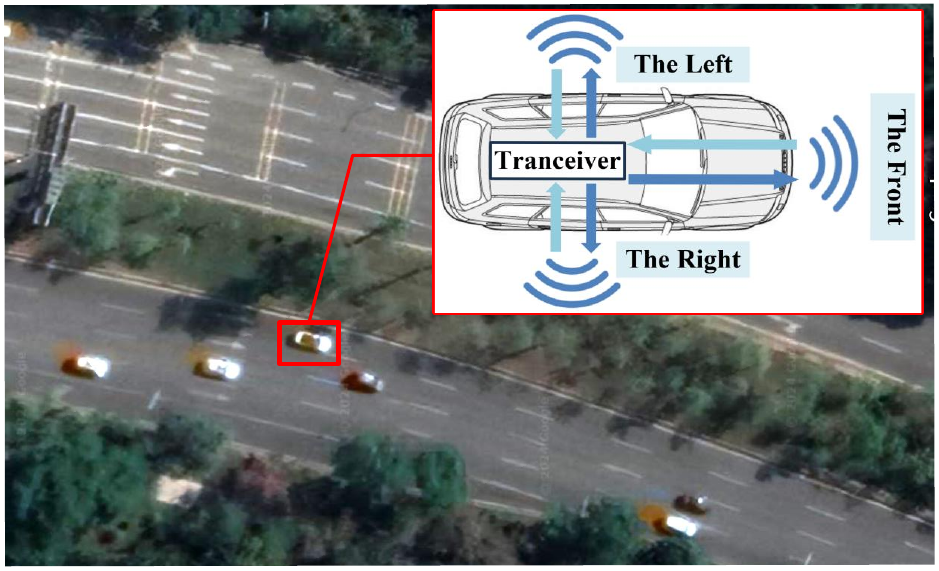}}
\subfigure[]{\includegraphics[width=1.1in]{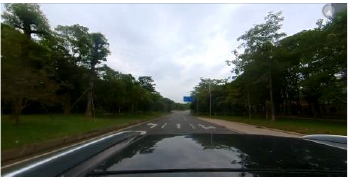}}
\subfigure[]{\includegraphics[width=1.1in]{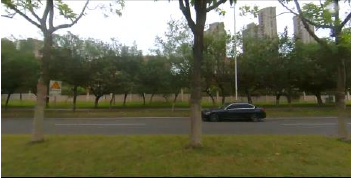}}
\subfigure[]{\includegraphics[width=1.1in]{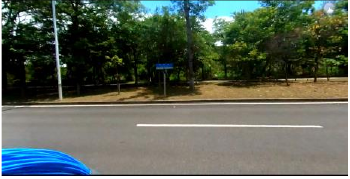}}
\caption{Measurement scenarios. (a) Measurement route and satellite photographs. (b) Sensing directions during the measurements. (c) Camera view from the front. (d) Camera view from the left. (e) Camera view from the right.}
\label{fig}
\end{figure}

The measurements were conducted in the Songshanhu district of Dongguan, located in Guangdong, China. The red line in Fig.3 (a) shows the trajectory diagram of vehicles for a single measurement route. The mean vehicle speed is 30 km/h and duration of measurements is generally between 20 and 25 minutes. From the satellite photographs in Fig. 3 (a), the measured streets are surrounded by buildings, trees, shrubs, etc., exhibiting varying environmental characteristics in different sensing directions. The sensing directions during the measurements are shown in Fig.3 (b), including front, left and right directions. The camera view from different directions are shown as Fig.3 (c)-(e). It is obvious that different directions have their own primary scatterers, leading to distinct environmental characteristics as follows:

$\bullet$ $\textit{Front}$: The main sensing scatterers include traffic lights in front of RX and preceding vehicles. Due to the forward extension of the road, the behavior of scatterers is mainly observed as being close to or far from the transceiver, resulting in a large dynamic range of distances for these scatterers from the transceiver. Besides, due to the high-speed movement, there aren't always sensing scatterers present in the front during the measurements.


$\bullet$ $\textit{Left}$: The main sensing scatterers include the median barriers, vehicles on the opposite road, and distant buildings. Due to the rapid movement of vehicles, these scatterers are often only perceptible for a short period of time. Besides, due to the distribution of these scatterers along the road, there are noticeable sensing scatterers for the majority of the measurement time.


$\bullet$ $\textit{Right}$: The main sensing scatterers include vehicles on the same-direction road and nearby buildings. Similar to the left direction, these scatterers often appear for only a short period of time. However, due to being on the same side of the road, vehicles traveling in the same direction may be perceptible for a longer duration. Additionally, the S-MPCs contributed by nearby buildings have higher powers than left.

Therefore, we statistically model the three main sensing directions separately in this paper, obtaining directional ISAC channel models.

\section{Data Processing}

\subsection{Power Delay Profile}

\begin{figure}[tbp]
\centering
\subfigure[]{\includegraphics[width=3in]{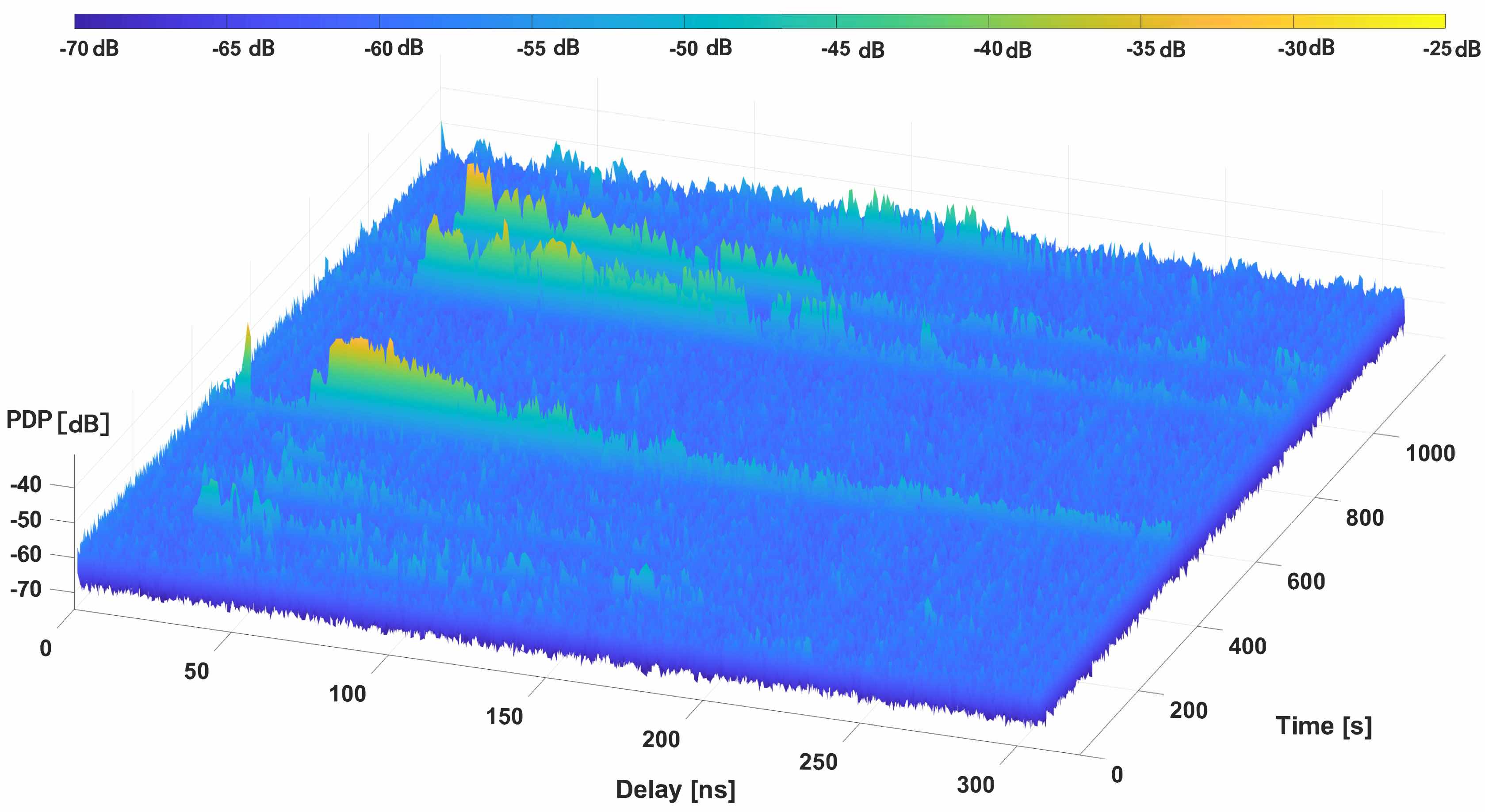}}
\subfigure[]{\includegraphics[width=3in]{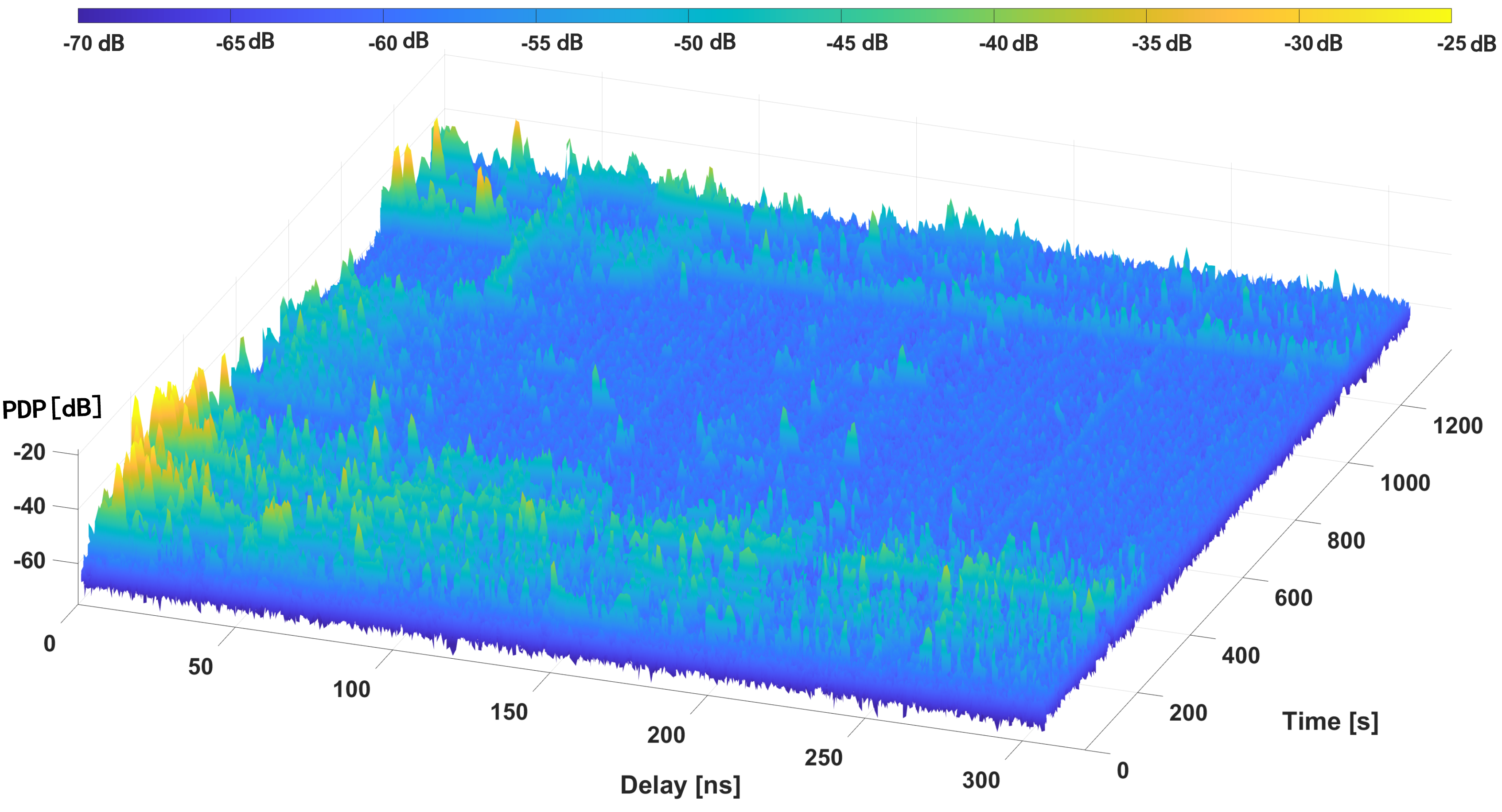}}
\subfigure[]{\includegraphics[width=3in]{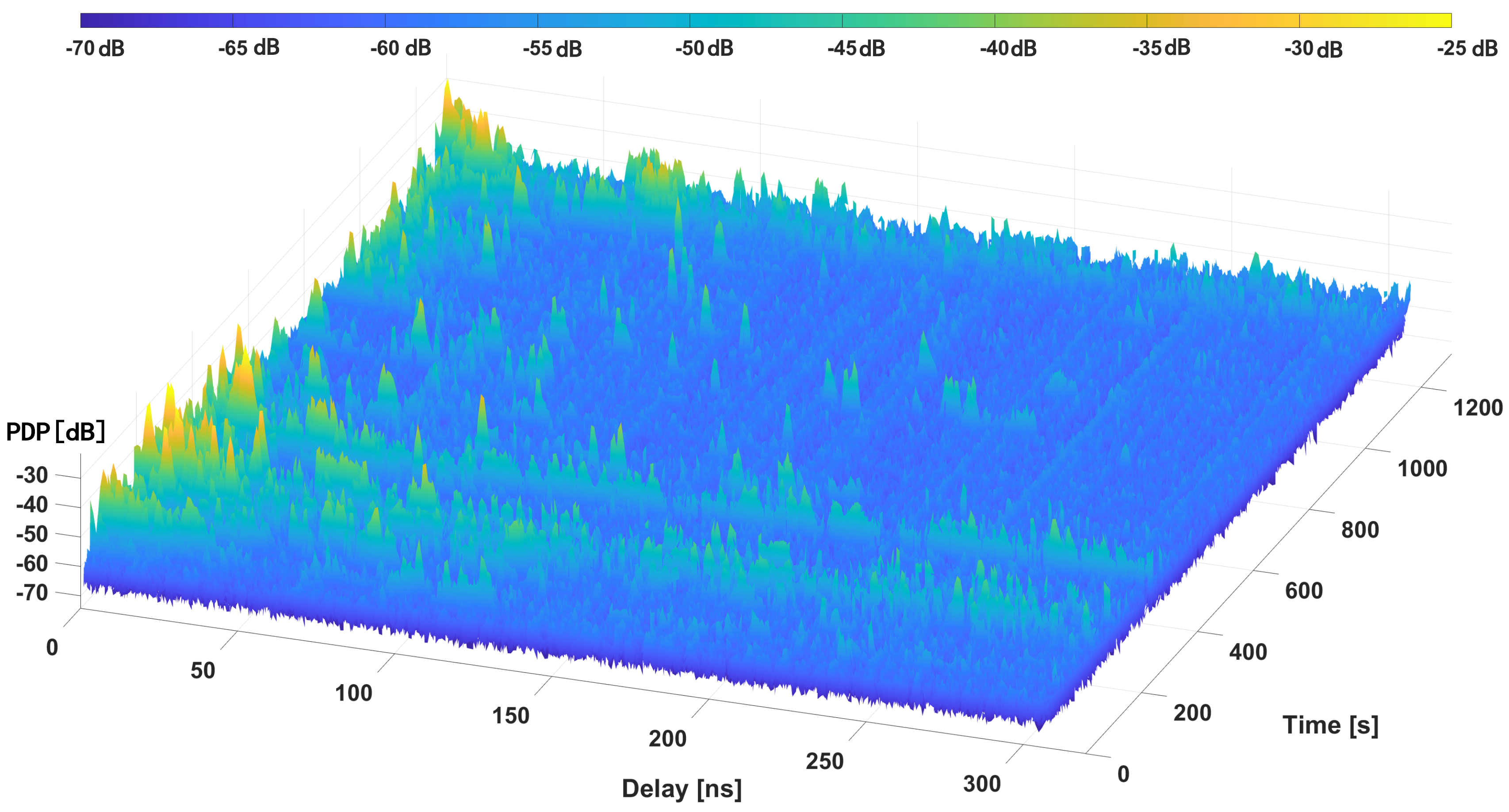}}
\caption{Dynamic PDPs of vehicular ISAC channels in different directions. (a) Front. (b) Left. (c) Right.}
\label{fig}
\end{figure}

During the measurements, due to the long driving route, unavoidable situations include waiting for traffic red lights. Therefore, prior to data processing, we eliminate intervals where the vehicle is static for waiting at traffic lights, based on synchronized video data, ensuring the continuous dynamics of the channel. Power delay profiles (PDPs) are widely used to reflect the received paths with propagation delays, which is calculated as follows:
\begin{equation}
P D P(t, \tau)=|h(t, \tau)|^2
\end{equation}
The PDPs of different directions are shown in fig. 4, which have different dynamic channel characteristics. It can be observed that the power in the left or right directions is higher than that in the front directions. This is attributed to scatterers, such as trees and buildings along the road, which result in strong and continuous echo signals, leading to higher energy in the left or right directions when the vehicle is in motion. Furthermore, the diversity of multipath in the left or right direction is significantly greater than that in the front directions, and tend to occur at low delays and span across a wider range in the time domain. Besides, for the front direction, multipaths exhibit sparsity in the time domain and a wider range of delays, which is attributed to the larger spatial depth of perceived targets, such as preceding vehicles, and there is a gap at low delays, which can be attributed to maintaining a safe driving distance.

To distinguish between S-MPCs and C-MPCs, we apply a sensing threshold for the initial preprocessing of measured PDPs, as illustrated in Fig. 5. The sensing threshold in this paper is set to be 6 dB above the noise floor. The S-MPCs have more power and are effective and positive  for sensing, while C-MPCs lack sufficient power to be perceived and are usually submerged in noise.

\begin{figure}[tbp]
\centering
\subfigure[]{\includegraphics[width=3in]{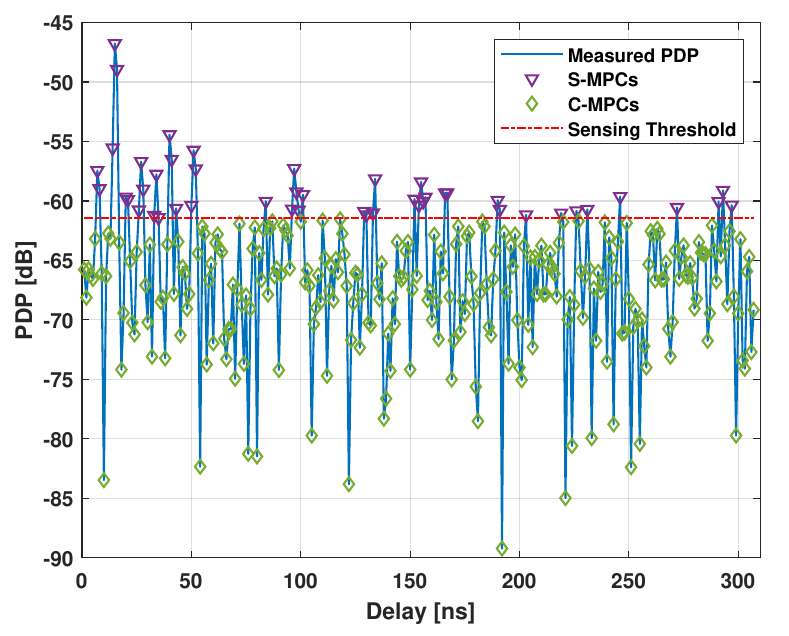}}
\caption{Distribution of S-MPCs and C-MPCs based on sensing threshold.}
\label{fig}
\end{figure}

\begin{figure}[tbp]
\centering
\subfigure[]{\includegraphics[width=1.1in]{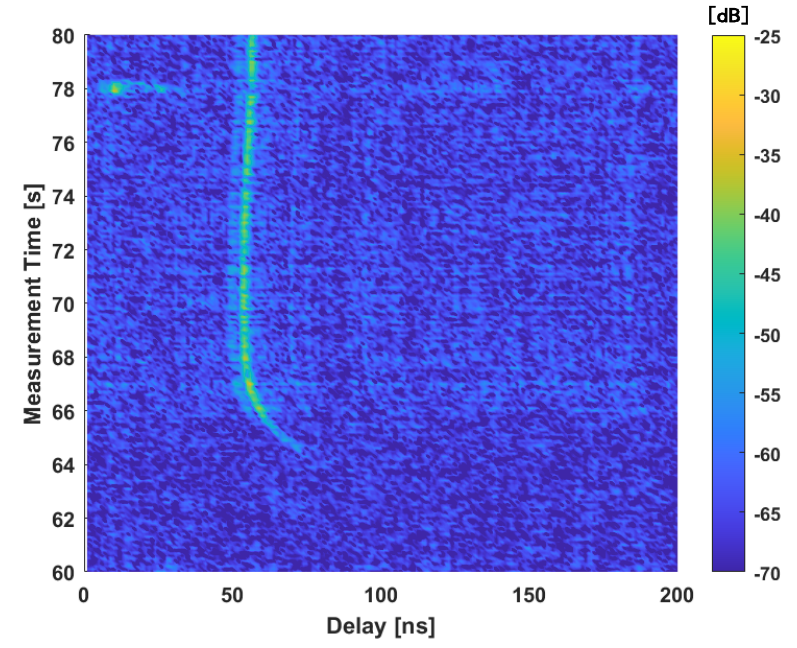}}
\subfigure[]{\includegraphics[width=1.09in]{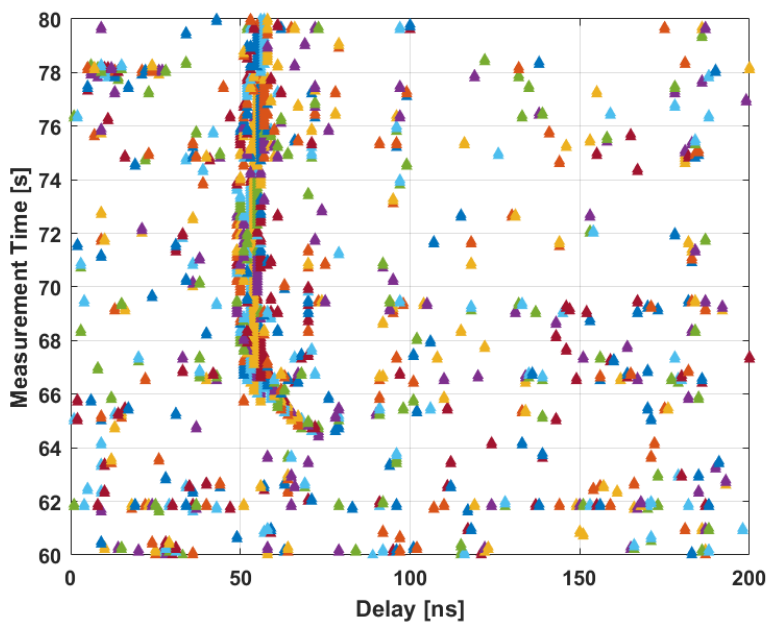}}
\subfigure[]{\includegraphics[width=1.1in]{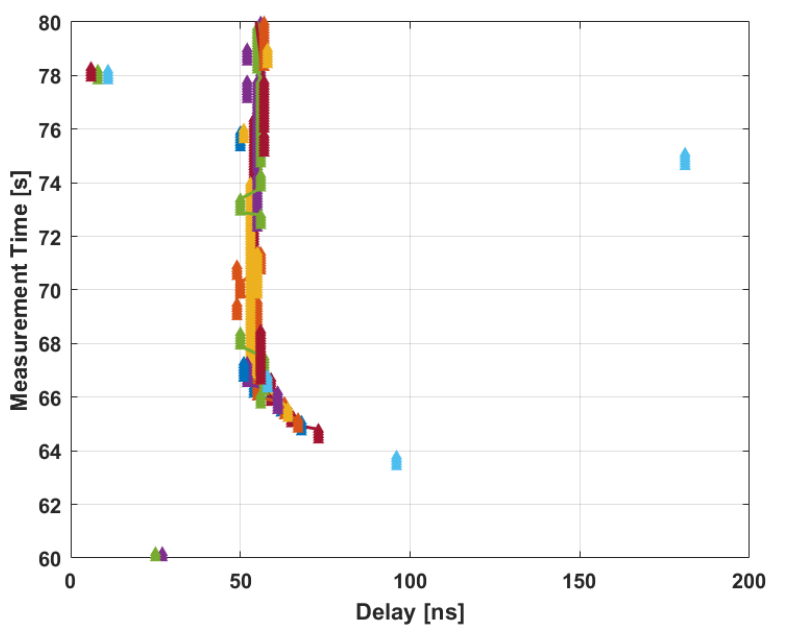}}
\caption{S-MPC tracking results. (a) Raw PDPs. (b) After sensing threshold. (c) After filtering and handover.}
\label{fig}
\end{figure}

\subsection{S-MPCs Modeling}

\subsubsection{S-MPC Tracking}

\begin{figure*}[tbp]
\centering
\subfigure[]{\includegraphics[width=2.3in]{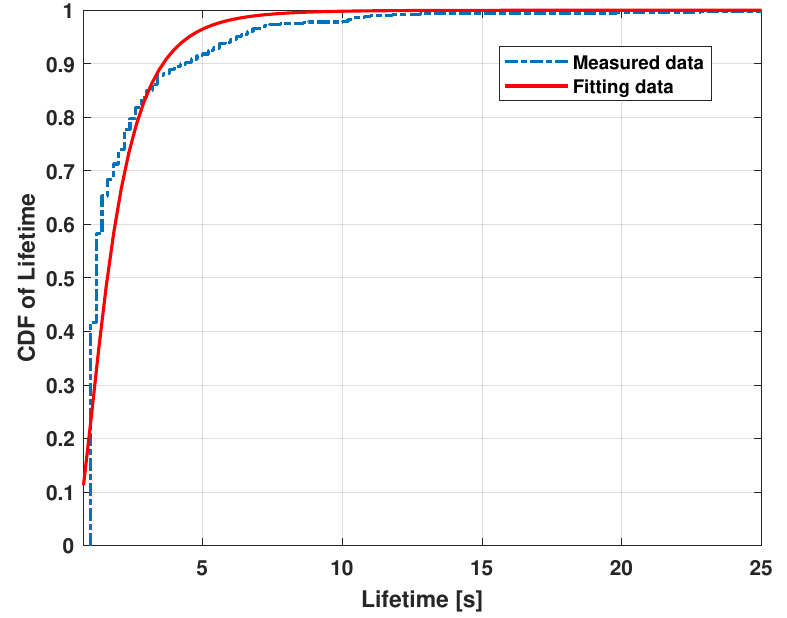}}
\subfigure[]{\includegraphics[width=2.3in]{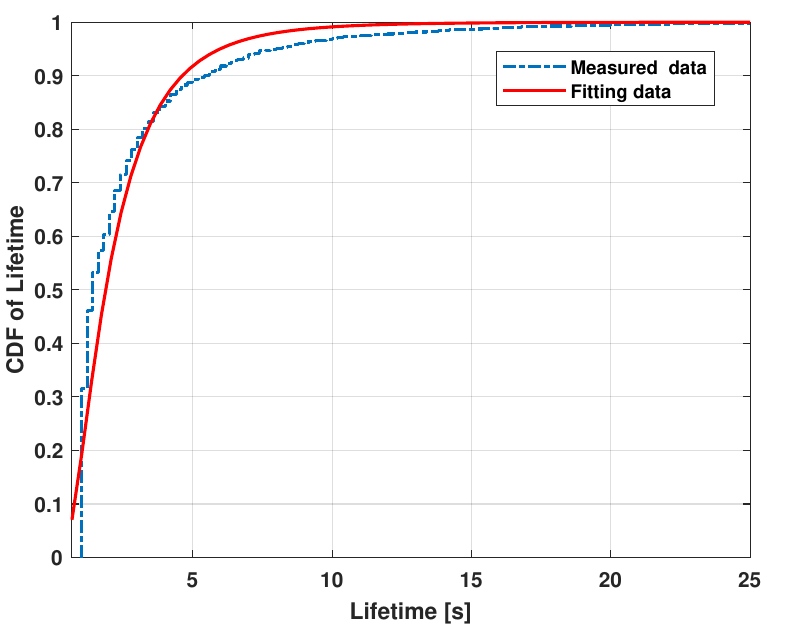}}
\subfigure[]{\includegraphics[width=2.3in]{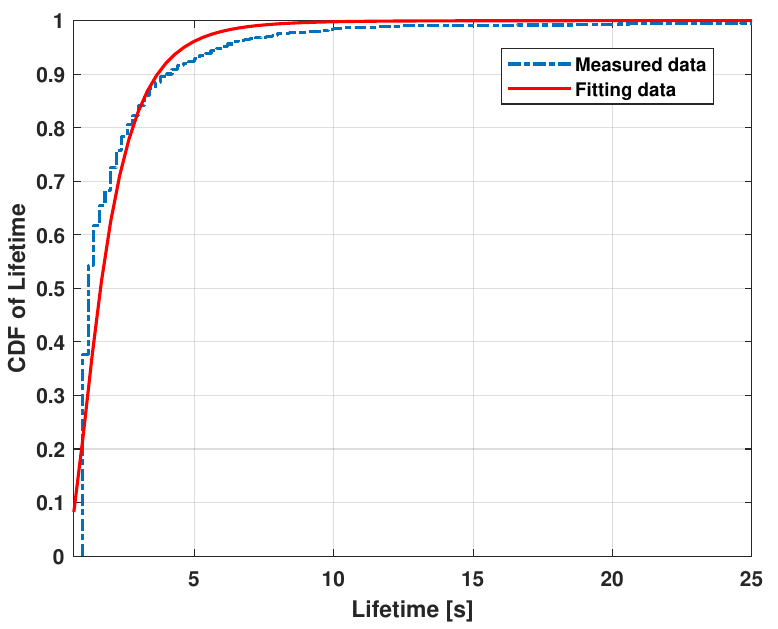}}
\caption{CDFs of lifetimes. (a) Front. (b) Left. (c) Right.}
\label{fig}
\end{figure*}

In practice, it is possible that the powers of some sudden clutters are higher than sensing threshold. Therefore, relying only on the sensing threshold is insufficient for extracting S-MPCs. Compared to C-MPCs, S-MPCs occur continuously over time. Thus, we use a MPC distance (MCD)-based tracking algorithm to extract S-MPCs more accurately and reasonably. MCD is used to measure the distance between two MPCs, where a smaller MCD indicates a higher degree of similarity between two MPCs. A heuristic normalized amplitude factor is introduced to enhance accuracy\cite{18}, and the calculation for MPCs $l_1$ and $l_2$ is as follows:
\begin{equation}
\begin{split}
&MCD= \\
&\left(\frac{1}{\Delta a_{\max }}\right)\left|\frac{{a_1\left(t_{i}\right)}}{{a_2\left(t_{i+1}\right)}}\right| \cdot \frac{\tau_{s t d}}{\left(\Delta \tau_{\max }\right)^2} \cdot\left|{\tau_1\left(t_i\right)}-{\tau_2\left(t_{i+1}\right)}\right|
\end{split}
\end{equation}
where
\begin{equation}
\Delta \tau_{\max }=\max \left({\tau_l}\right)-\min \left({\tau_l}\right)
\end{equation}
\begin{equation}
\Delta a_{\max }=\frac{\max \left({a_l}\right)}{\min \left({a_l}\right)}
\end{equation}
and $a_1$ and $a_2$ are amplitudes of MPCs $l_1$ and $l_2$ respectively, $\tau_1$ and $\tau_2$ are delays of MPCs $l_1$ and $l_2$ respectively, $a_l$, $\tau_l$ and $\tau_{std}$ are amplitudes, delays and a standard devistion of delays for all MPCs at time $t_i$ and $t_{i+1}$. The $|\cdot|$ is the absolute value. 

In the proposed MCD-based tracking algorithm, three specific thresholds are defined: the matching threshold $\epsilon_m$, filtering threshold $\epsilon_f$, and handover threshold $\epsilon_h$. The $\epsilon_m$ is used to match two MPCs based on the similarity of multipaths, i.e., MCD. The $\epsilon_f$ is used to filter out sudden multipaths, which may be caused by clutters. The $\epsilon_h$ is used to hand over non-successive multipaths that are closely spaced in both time and delay domains (distinguished as $\epsilon_{h_{t}}$ and $\epsilon_{h_{d}}$), resulting from actual measurements and being unavoidable. The proposed MCD-based tracking algorithm is described as follows:

\begin{itemize}
\item[a)]
 An MCD matrix $\textbf{D}$, meaning similarity between any S-MPC at time $t_i$ and any S-MPC at time $t_{i+1}$, is calculated with dimension $L(t_i)\times L(t_{i+1})$.
\end{itemize}
\begin{itemize}
\item[b)]
A unique S-MPC ID is assigned to match the same S-MPC, which is determined based on conditions as follows:
\begin{equation}
\left\{\begin{array}{c}
\mathrm{\textbf{D}}_{u, v} \leq \epsilon_m \\
u=\underset{u}{\arg \min } (\mathrm{\textbf{D}}_{u \in L\left(t_i\right), v}) \\
v=\underset{v}{\arg \min } (\mathrm{\textbf{D}}_{u, v \in L\left(t_{i+1}\right)})
\end{array}\right.
\end{equation}
If equation (9) is satisfied, the $u$th S-MPC at time $t_i$ and the $v$th S-MPC at time $t_{i+1}$ are considered to be the same S-MPC. Repeat this step to examine all S-MPCs at time $t_i$ and time $t_{i+1}$.
\end{itemize}
\begin{itemize}
\item[c)]
Repeat steps 1 and 2 to calculate MCD between any S-MPC at time $t_{i+1}$ and any S-MPC at time $t_{i+2}$, and obtain new S-MPC IDs. If the $w$th S-MPC at time $t_{i+2}$ is found to match the $v$th S-MPC at time $t_{i+1}$, the $w$th S-MPC at time $t_{i+2}$ inherits the S-MPC ID from the $v$th S-MPC at time $t_{i+1}$, and so forth, obtaining S-MPC IDs for the whole measurements.
\end{itemize}
\begin{itemize}
\item[d)]
Sudden multipaths are filtered out based on filtering threshold $\epsilon_f$. If the length of S-MPC is less than $\epsilon_f$, this multipath is removed from the set of S-MPCs.
\end{itemize}
\begin{itemize}
\item[e)]
Hand over S-MPCs that are close to each other. If the time between S-MPCs $l_u$ and $l_v$ is less than $\epsilon_{h_{t}}$, and the delay between S-MPCs $l_u$ and $l_v$ is less than $\epsilon_{h_{d}}$, the S-MPC $l_u$ inherits the S-MPC ID from the S-MPC $l_v$. Repeat this step to hand over all potential S-MPCs.
\end{itemize}

Through the preceding algorithm, all S-MPCs are extracted and tracked over time, allowing the lifetime and evolution of each S-MPC to be obtained. Based on manual validation, we set $\epsilon_m=0.3$, $\epsilon_f=4$, $\epsilon_{h_t}=5$ and $\epsilon_{h_d}=10$, respectively. 574, 2885, and 1625 S-MPCs are extracted for the front, left, and right directions, respectively. This indicates that the number of S-MPCs is the highest for the left, while it is the least for the front, which is consistent with the observation from Fig. 4.

An example plot of S-MPC tracking results is shown in Fig. 6, including raw PDPs, S-MPCs results after sensing threshold and S-MPCs results after filtering and handover, respectively. Different color means different extracted S-MPCs in Fig. 6(b) and (c). It can be observed that effective sensing multipaths in raw PDPs are tracked well and reasonably based on S-MPC tracking results, and most of the clutter is eliminated. Besides, S-MPCs with similar time and delay characteristics are clustered, which needs to be considered in the following dynamic modeling approach.

\subsubsection{Lifetime}

Lifetime is the duration of existence for S-MPC, measured from the first observed snapshot to the last observed snapshot. Based on the tracking results, we model the distribution of all S-MPCs in different directions. Fig.7 shows the cumulative density function (CDF) fit to the measurements, and the log-normal distribution are plotted for comparison. Although variations in the scatterers distribution and environmental characteristics, the lifetime of S-MPCs in different directions are generally consistent, with 90$\%$ of lifetimes falling within 5 seconds. The lifetime in the left is slightly longer than that in other directions, which is caused by noticeable sensing scatterers. Finally, we model lifetime as a log-normal distribution with mean value $\mu_{T}$ and standard devistion $\sigma_{T}$ as follows, which are summarized in Table $\textrm{IV}$.
\begin{equation}
f(x|\mu_{T}, \sigma_{T}) = \frac{1}{x \sigma_{T} \sqrt{2\pi}} \exp\left(-\frac{(\ln x - \mu_{T})^2}{2\sigma_{T}^2}\right)
\end{equation}

\subsubsection{Number of New S-MPCs}

As mentioned in section $\textrm{II}$, the number of all S-MPCs $L(t_i)$ at time $t_i$ consists of two parts: old S-MPCs that appeared before time $t_i$, denoted as $L(t_{j,i})$, and new S-MPCs that appeared at time $t_i$, denoted as $L(t_{i,i})$, where $0 < j < i$. Note that $L(t_{j,i})$ is determined by the total number $L(t_{j})$ at time $t_j$ and their lifetimes, therefore, only the number of new S-MPCs at each time instant is needed to model to simulate dynamic channel. Based on the tracking results, we counted the number of new S-MPCs in each snapshot. Due to the discreteness and sparsity of S-MPCs, we do not establish a distribution fitting model for parameters; instead, we adopt their number probabilities as the distribution model, as shown in Table $\textrm{II}$. It is found that in the majority of snapshots, there are no new S-MPCs appearing (the number of new S-MPCs is zero), especially in front direction, with a 96.03$\%$ probability of no new S-MPCs. Besides, not many new S-MPCs appear at a single time instant simultaneously (maximum number of new S-MPCs is five). This indicates that S-MPCs in the vehicular ISAC channel are sparse, and this sparsity is more pronounced in front direction compared to other directions, which is reasonable according to actual measurements.

\begin{table}[]
\centering
\caption{New S-MPCs number probability.}
\begin{tabular}{ccccccc}
\hline
New S-MPCs number & 0     & 1     & 2    & 3    & 4     & 5     \\ \hline
Front (\%)           & 96.03 & 3.47  & 0.39 & 0.09 & 0.01 & 0.01 \\
Left (\%)            & 82.58 & 14.84 & 2.09 & 0.33 & 0.11  & 0.01 \\
Right (\%)           & 89.42 & 9.07  & 1.22 & 0.23 & 0.05 & 0.01 \\ \hline
\end{tabular}
\end{table}

\begin{figure}[tbp]
\centering
\subfigure[]{\includegraphics[width=1.6in]{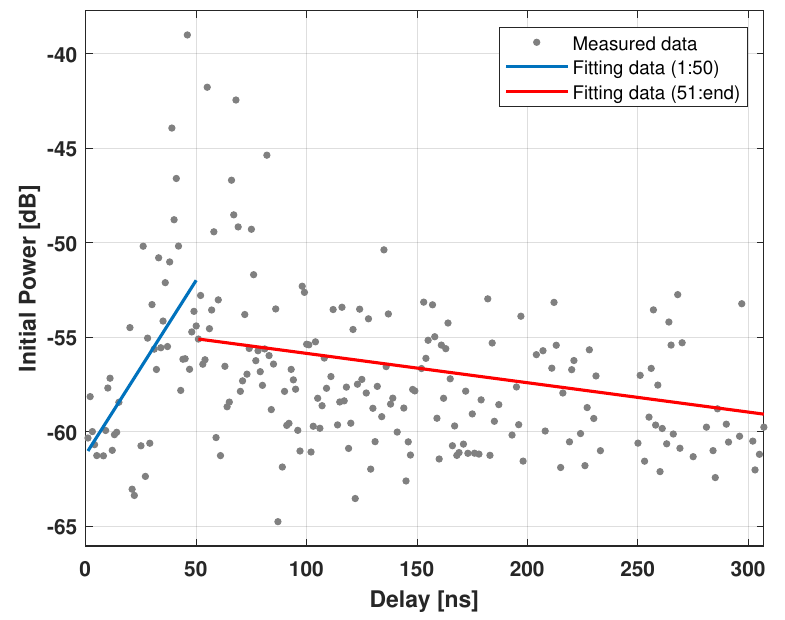}}
\subfigure[]{\includegraphics[width=1.6in]{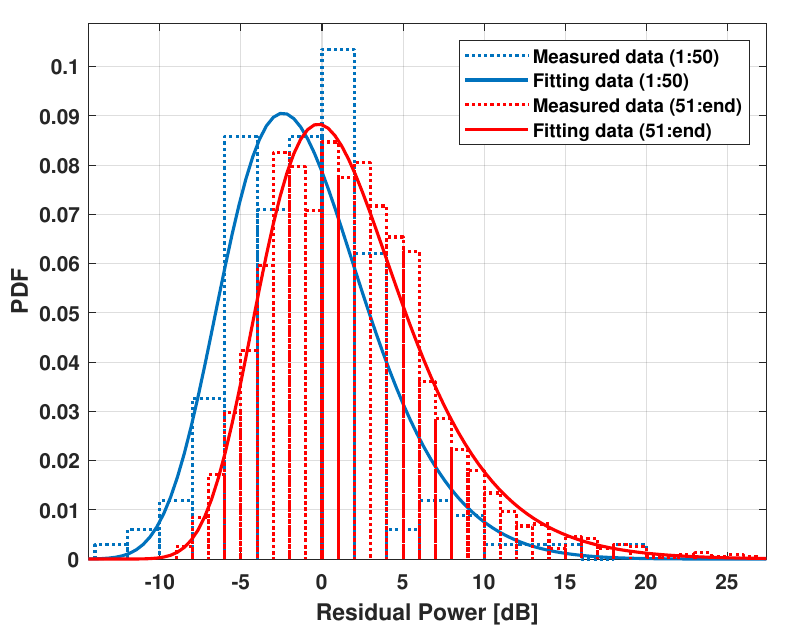}}
\subfigure[]{\includegraphics[width=1.6in]{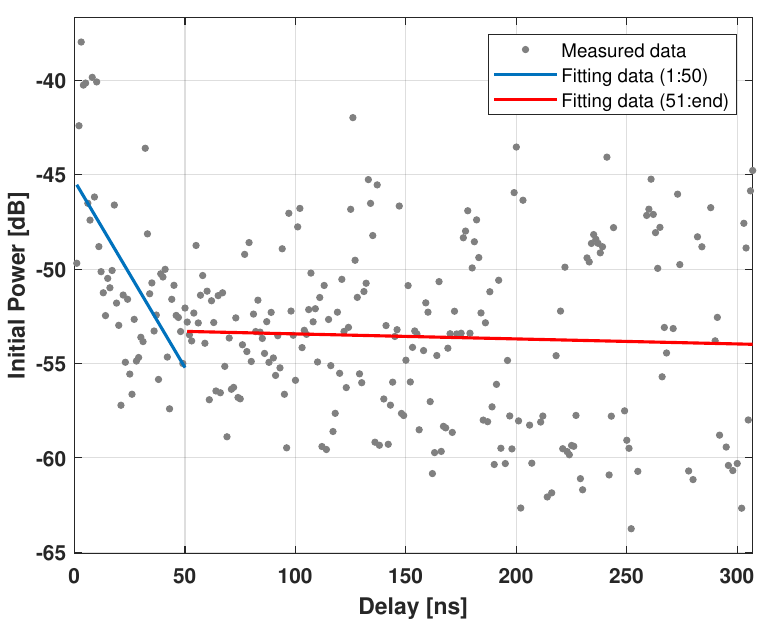}}
\subfigure[]{\includegraphics[width=1.6in]{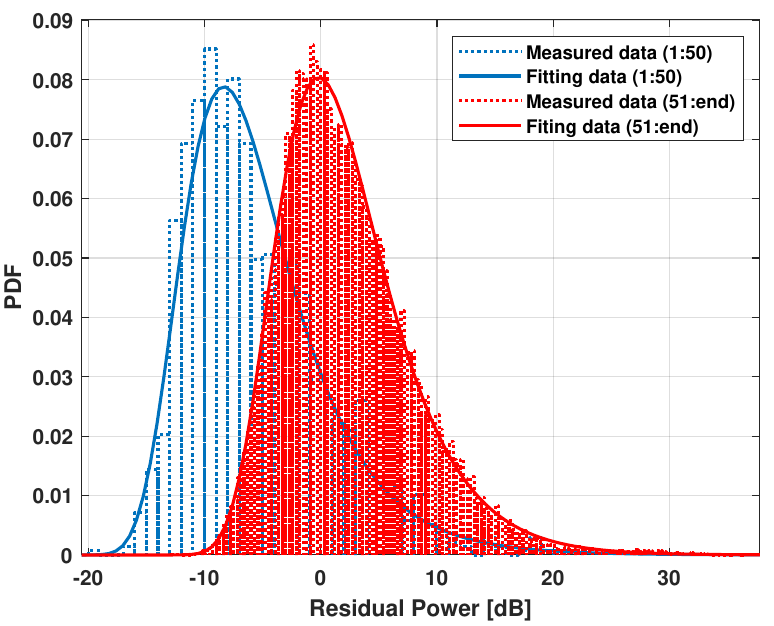}}
\subfigure[]{\includegraphics[width=1.6in]{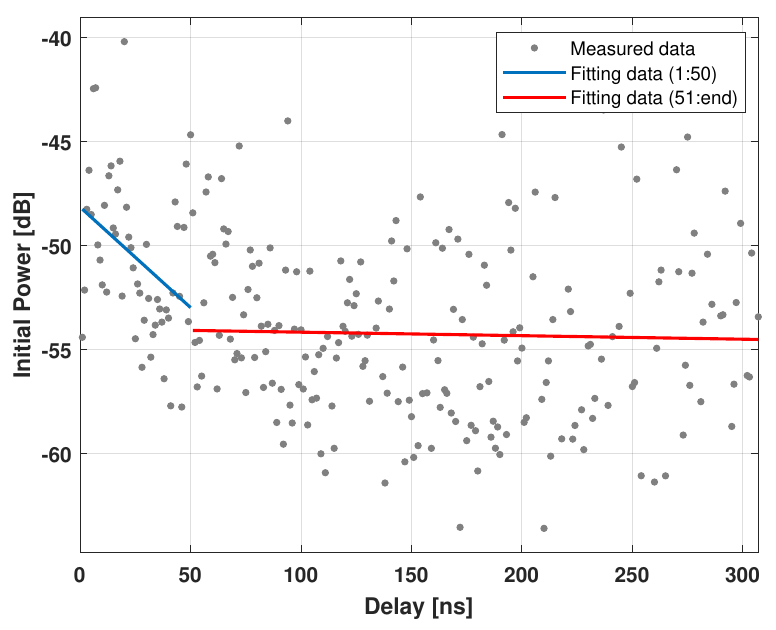}}
\subfigure[]{\includegraphics[width=1.6in]{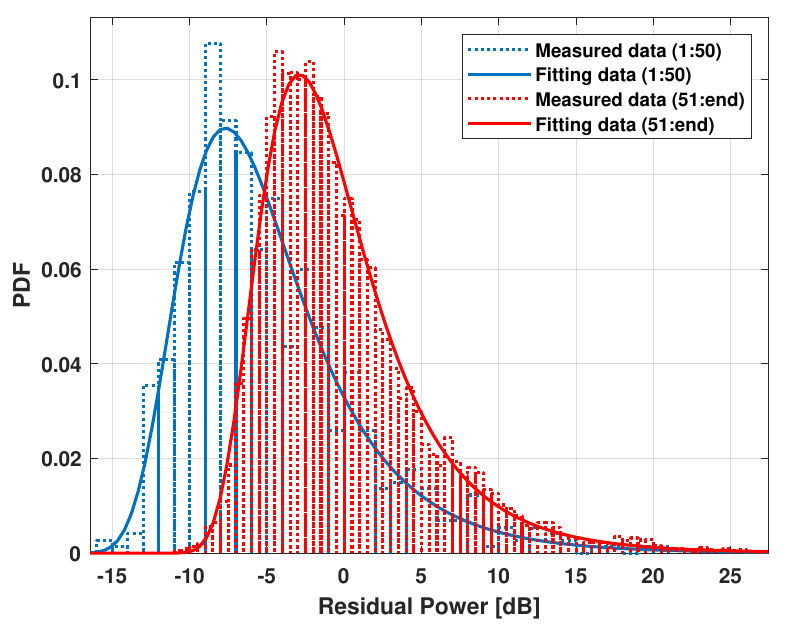}}
\caption{Initial power in different directions. (a) Dual-slope fitting (Front). (b) Residual power (Front). (c) Dual-slope fitting (Left). (d) Residual power (Left). (e) Dual-slope fitting (Right). (f) Residual power (Right)}
\label{fig}
\end{figure}

\subsubsection{Initial Power and Delay}

We focus on characterizing the initial power distribution and initial delay position distribution of new S-MPCs. In fact, the power and delay of old S-MPCs can be effectively modeled through the evolution model within lifetime, together with initial power and delay when those old S-MPCs were initially generated. 

Generally, power is negatively correlated with delay, i.e., the larger delay, indicating a longer propagation path, the lower power of multipath. Therefore, we model initial power as a dual-slope function of delay, together with a residual power distribution. We set 50 ns as the breakpoint of dual-slope function, because the corresponding sensing distance for 50 ns is 7.5 meters, covering the majority of the driving roads. S-MPCs with delays exceeding 50 ns often originate from scatterers outside the road, such as trees and buildings, and typically have lower power. To have a good fit to measurements, all power is modeled in decibel scale in this paper. Fig. 8(a), (c) and (e) show the initial power in different directions, and the dual-slope functions are plotted for comparison. Note that there is a gap at low delays in front direction from Fig. 4(a), resulting in power being positively correlated with delay in this area. We model initial power as a dual-slope function with parameters $p$ and $q$, as follows:
\begin{equation}
a_l(t,\tau)= \begin{cases}p_1 \cdot \tau[\mathrm{ns}]+q_1, & 0 \leq \tau<50 ns \\ p_2 \cdot \tau[\mathrm{ns}]+q_2, & 50 ns \leq \tau \end{cases}
\end{equation}
Actually, the relationship between power and delay is not a deterministic function, thus, we also model the residual power distribution. Fig. 8(b), (d) and (e) shows the probability density function (PDF) of residual power. Due to different number of tracked S-MPCs, the number of samples in Fig. 8(d) is higher than others. We model residual power as a Generalized extreme value distribution with shape parameter $\xi_r$, scale parameter $\sigma_r$, and location parameter $\mu_r$, as follows:
\begin{equation}
f(x|\mu_r, \sigma_r, \xi_r) = \frac{1}{\sigma_r} t(x)^{\xi_r+1} e^{-t(x)}
\end{equation}
where
\begin{equation}
t(x)=\left[1+\xi_r\left(\frac{x-\mu_r}{\sigma_r}\right)\right]^{-\frac{1}{\xi_r}}
\end{equation}
Finally, the initial power of S-MPCs is obtained by adding the residual power to delay-based power. The specific fitting values are summarized in Table $\textrm{IV}$.

Fig. 9 shows the CDF of initial delay of S-MPCs in different directions, and the Gamma distributions are plotted for comparison. It can be observed that the initial delay in front direction has a wider range compared to other directions. For instance, the initial delay of 50$\%$ of S-MPCs in front direction can reach to 100 ns, whereas in the left and right directions, it is less than 50 ns. This is consistent with the characteristics mentioned in Section III(B), which demonstrate a large dynamic range of S-MPCs in the delay domain. We model initial delay as a Gamma distribution with the shape parameter $\alpha_d$ and the scale parameter $\beta_d$, as follows:
\begin{equation}
f(x|\alpha_d, \beta_d) = \frac{{\beta_d}^{\alpha_d} x^{\alpha_d-1} e^{-\beta_d x}}{\Gamma(\alpha_d)}
\end{equation}
where $\Gamma(\cdot)$ is the Gamma function. The specific fitting values are summarized in Table $\textrm{IV}$.

\begin{figure}[tbp]
\centering
\subfigure[]{\includegraphics[width=1.6in]{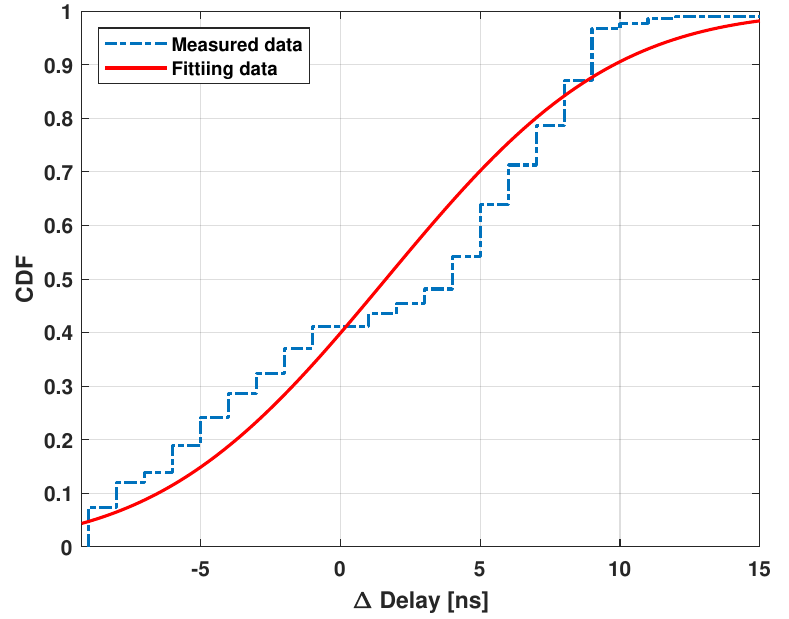}}
\subfigure[]{\includegraphics[width=1.6in]{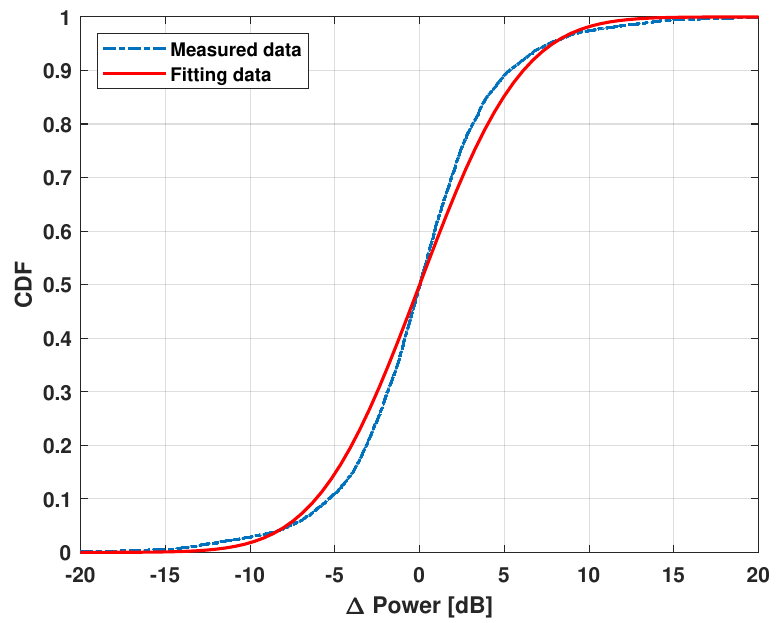}}
\subfigure[]{\includegraphics[width=1.6in]{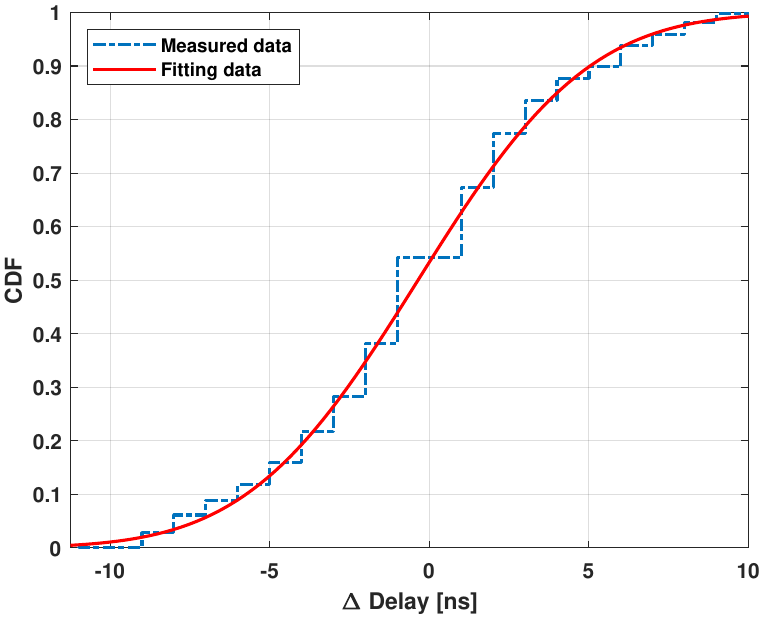}}
\subfigure[]{\includegraphics[width=1.6in]{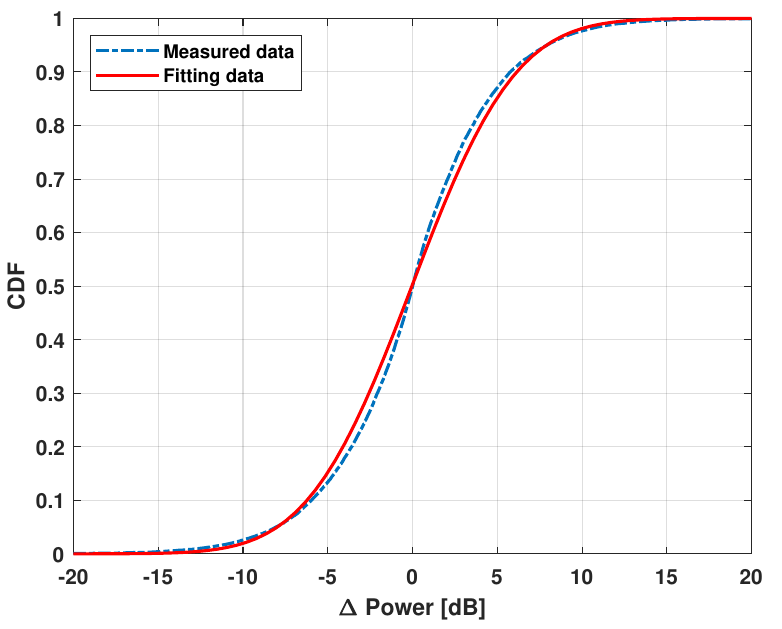}}
\subfigure[]{\includegraphics[width=1.6in]{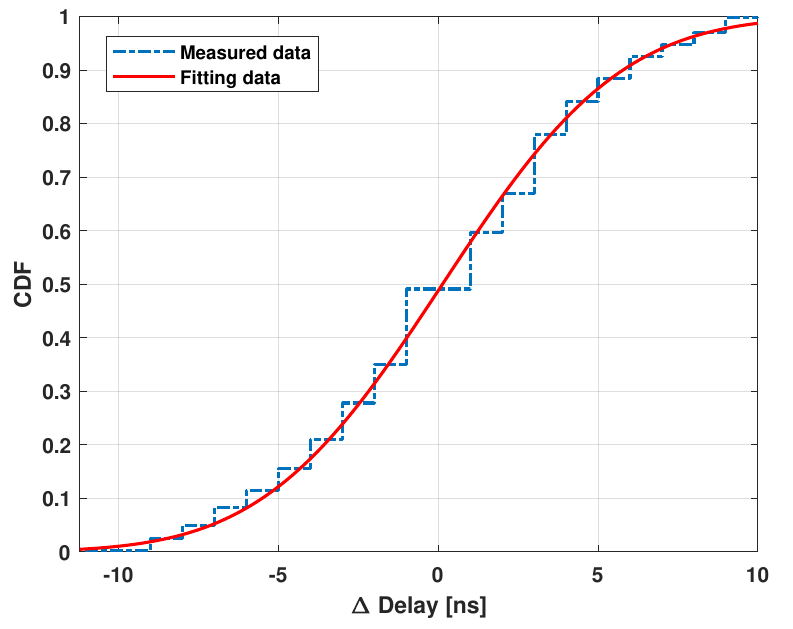}}
\subfigure[]{\includegraphics[width=1.6in]{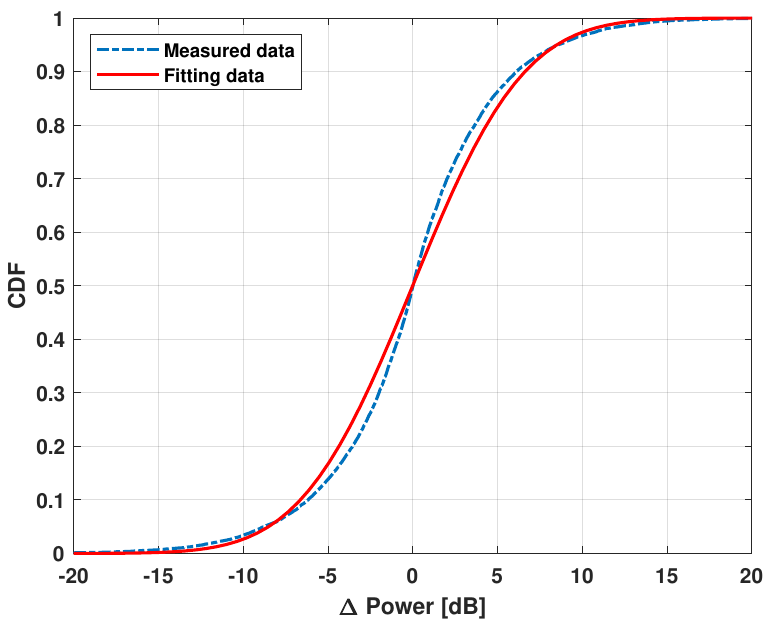}}
\caption{Evolution over successive time. (a) Delay evolution (Front). (b) Power evolution (Front). (c) Delay evolution (Left). (d) Power evolution (Left). (e) Delay evolution (Right). (f) Power evolution (Right).}
\label{fig}
\end{figure}

\begin{figure}[tbp]
\centering
\subfigure[]{\includegraphics[width=3.1in]{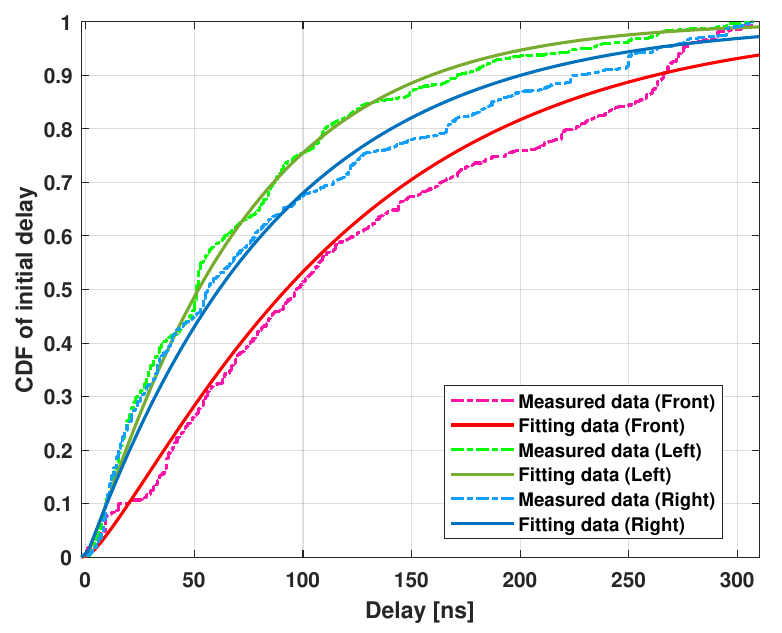}}
\caption{CDFs of initial delay in different directions.}
\label{fig}
\end{figure}

\subsubsection{Evolution Over Successive Time}
In this subsection, we model variations of power and delay of each S-MPCs within its lifetime. Since the initial power and delay of S-MPCs have been modeled, it is only necessary to model the distribution of variations over time to characterize dynamic evolution within lifetime. In our processing, we found that the majority of snapshots exhibit unchanged S-MPCs in the delay domain. Therefore, we consider using a threshold to determine whether delays of S-MPCs undergo dynamic evolution, denoted as $k_d$. Fig. 10 shows the CDF of variations of delay and power over successive time, and the Normal distributions are plotted for comparison. For powers, there is almost no difference among the three directions. For delays, there is a larger variation in front direction, consistent with the observation from Fig. 4, where multipaths exhibit a large dynamic range in the delay domain. We model evolution over successive time as a Normal distribution with the mean value $\mu_e$ and standard deviation $\sigma_e$, as follows, which are summarized in Table $\textrm{IV}$.
\begin{equation}
f(x|\mu_e, \sigma_e) = \frac{1}{\sigma_e \sqrt{2\pi}} \exp\left(-\frac{(x - \mu_e)^2}{2{\sigma_e}^2}\right)
\end{equation}

\begin{figure}[tbp]
\centering
\subfigure[]{\includegraphics[width=1.6in]{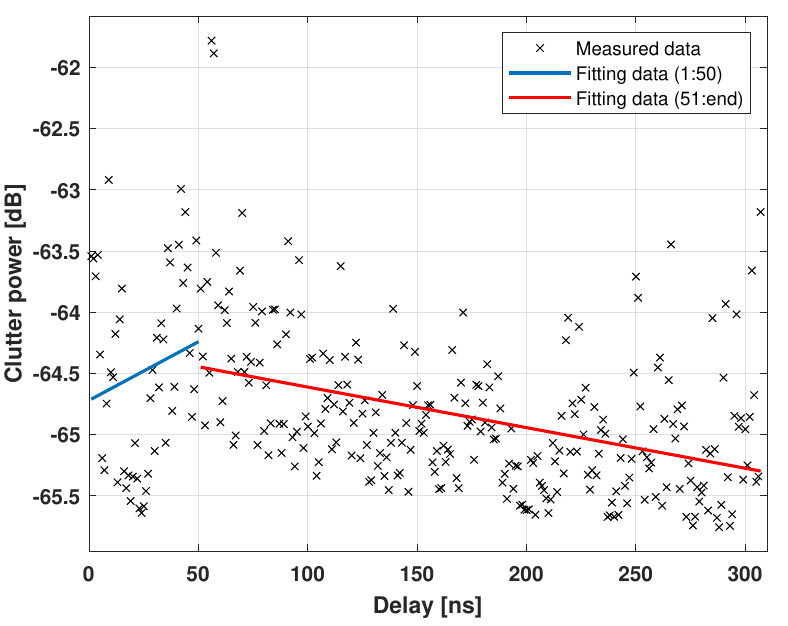}}
\subfigure[]{\includegraphics[width=1.6in]{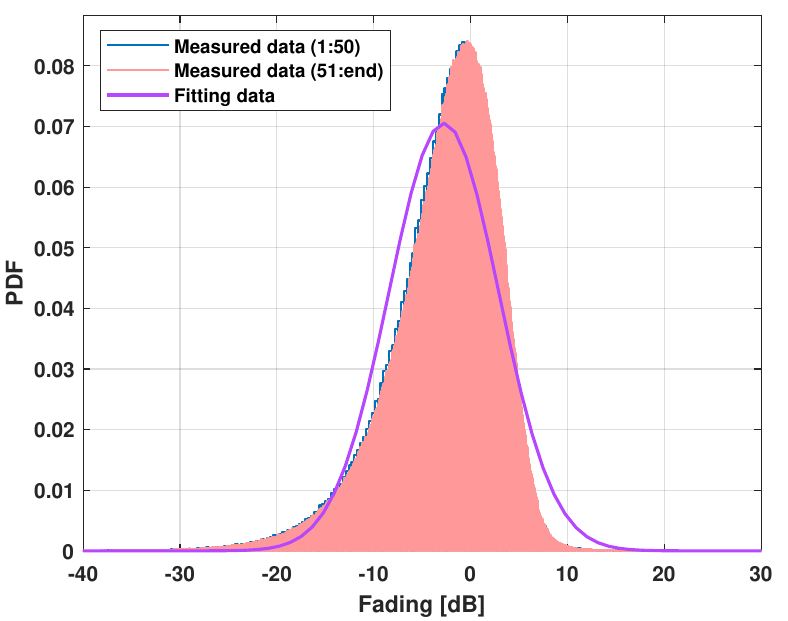}}
\subfigure[]{\includegraphics[width=1.6in]{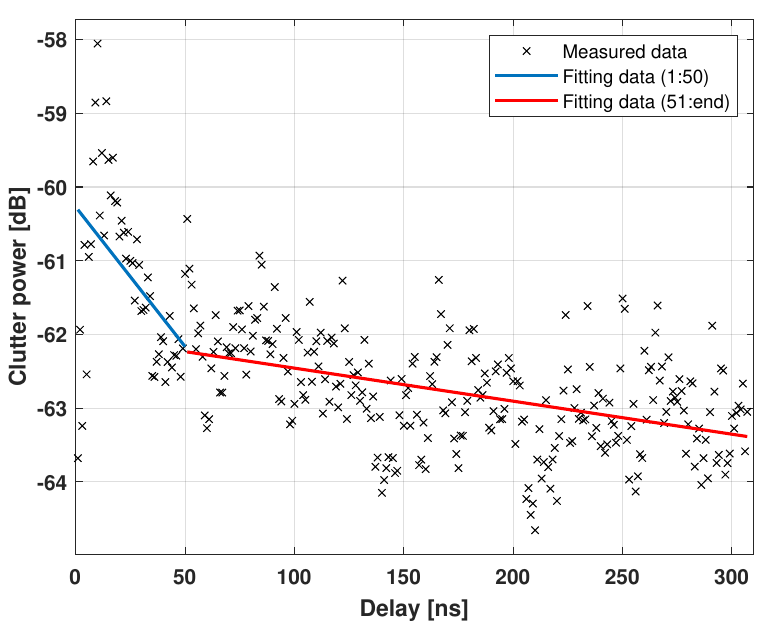}}
\subfigure[]{\includegraphics[width=1.6in]{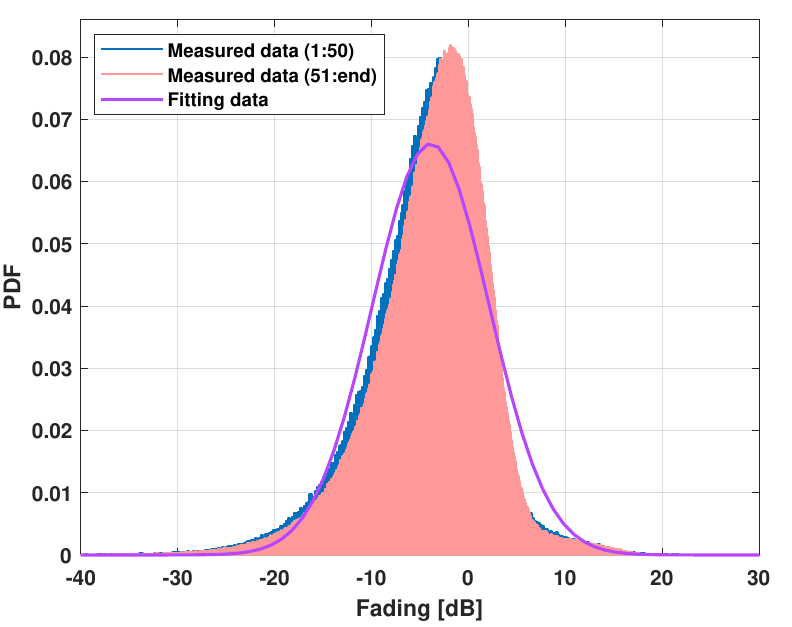}}
\subfigure[]{\includegraphics[width=1.6in]{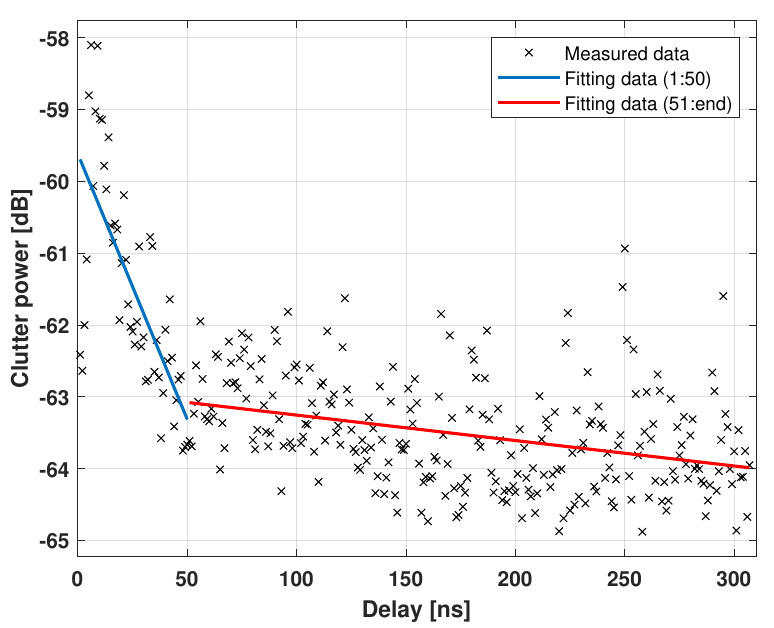}}
\subfigure[]{\includegraphics[width=1.6in]{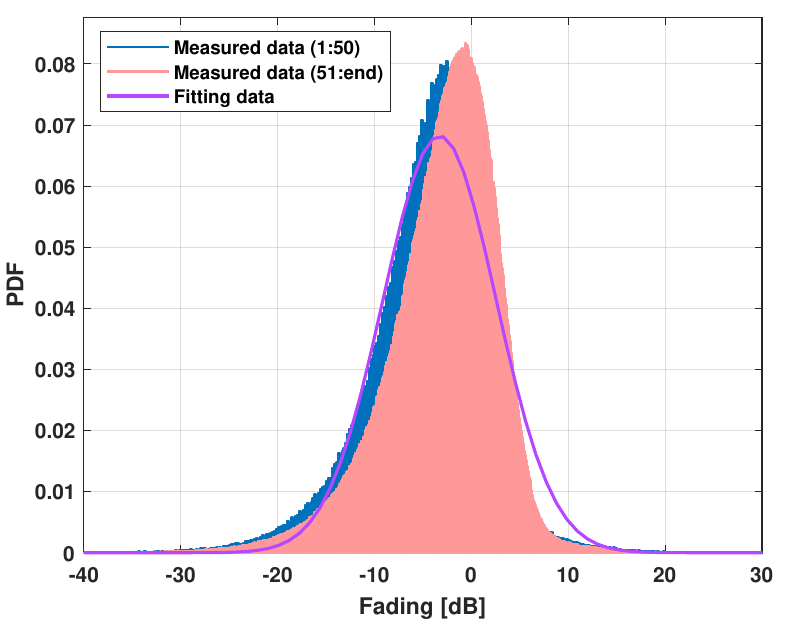 }}
\caption{Clutter power in different directions. (a) Dual-slope fitting (Front). (b) Fading (Front). (c) Dual-slope fitting (Left). (d) Fading (Left). (e) Dual-slope fitting (Right). (f) Fading (Right)}
\label{fig}
\end{figure}

\subsubsection{Clustering}

Lots of channel measurements have shown that MPCs are generally distributed in groups (clusters), which is suitable for vehicular ISAC channels. Based on the tracking results, we counted the number of S-MPCs within a cluster in each snapshot. Similar to the number of new S-MPCs, we consider probabilities of different numbers as the distribution model, as illustrated in Table III, rather than establishing a distribution fitting model. It can be observed that in more than 90$\%$ of clusters in any directions, the number of S-MPCs within a cluster does not exceed 4. For any newly generated S-MPCs, they tend to cluster with other S-MPCs before being assigned across other delay positions. Specifically, determine whether the surroundings of the currently existing S-MPCs satisfy the expected number. If they do not, the new S-MPC is adjustably assigned to be around the currently existing ones.

\begin{table}[]
\centering
\caption{Number of S-MPCs within a cluster.}
\begin{tabular}{cccccccc}
\hline
Number & 1     & 2     & 3     & 4    & 5    & 6    & 7    \\ \hline
Front (\%)     & 35.99 & 35.75 & 11.92 & 6.91 & 4.22 & 2.21 & 0.95 \\
Left (\%)      & 28.73 & 43.21 & 16.35 & 6.88 & 2.87 & 1.12 & 0.41 \\
Right (\%)     & 31.67 & 40.53 & 14.54 & 6.83 & 3.18 & 0.50 & 0.38 \\ \hline
\end{tabular}
\end{table}

\subsection{C-MPCs Modeling}

Since C-MPCs are randomly distributed across the entire delay domain and lack dynamic characteristics such as lifetime, time evolution, clustering and so on, we mainly focus on characterizing the power distribution for C-MPCs modeling. Similar to the initial power of S-MPCs, power of C-MPCs is negatively correlated with delay, and we model clutter power decay as a dual-slope function of delay, together with a small-scale fading power distribution. In order to keep consistency, 50 ns is still considered as the breakpoint of dual-slope function. Fig. 11(a), (c) and (e) show the clutter power decay in different directions, and the dual-slope functions are plotted for comparison. Compared to initial power of S-MPCs, clutter power is lower and exhibits a smaller range of variation. The fitting model is similar to equation (11), where parameters are replaced by $a$ and $b$. Fig. 11(b), (d) and (e) shows the PDF of small-scale fading of C-MPCs, which exhibits random variations in delay domain. It can be observed that the distribution of fading in [0-50] ns is similar to that of [50-end] ns in any directions. Therefore, we no longer segment the fading into two parts for modeling. Besides, in order to mitigate the impact of extreme C-MPCs, we model fading as a normal distribution with the mean value $\mu_c$ and standard deviation $\sigma_c$, consistent with equation (15).

\section{Model Implementation}
\subsection{Implementation}


\begin{figure}[tbp]
\centering
\subfigure[]{\includegraphics[width=2.5in]{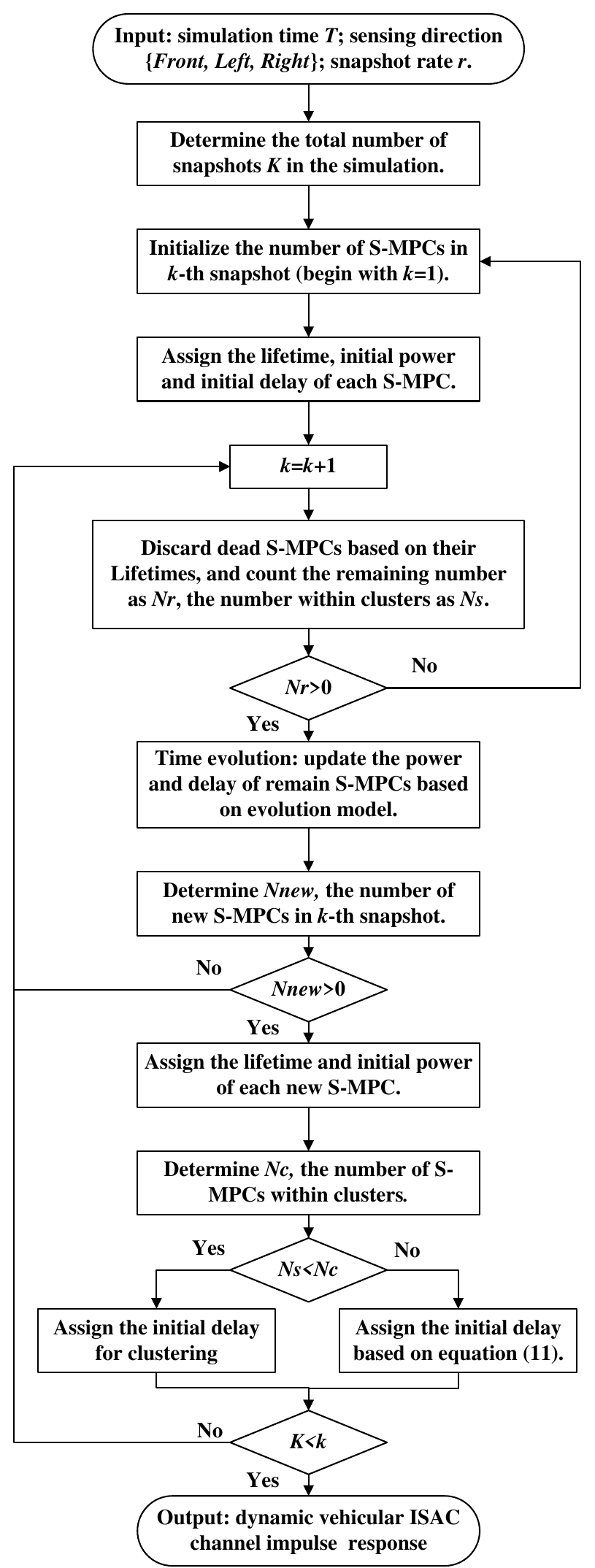}}
\caption{Flowchart of implementation steps.}
\label{fig}
\end{figure}

\begin{table}[]
\centering
\caption{The summary of statistic parameters for dynamic vehicular ISAC channels.}
\scalebox{0.8}{
\begin{tabular}{ccc|cccc}
\hline
\multicolumn{3}{c|}{Modeling}                                                                                                          & \multicolumn{1}{c|}{Parameter} & \multicolumn{1}{c|}{Front}   & \multicolumn{1}{c|}{Left}    & Right   \\ \hline
\multicolumn{1}{c|}{\multirow{21}{*}{S-MPCs}} & \multicolumn{2}{c|}{\multirow{2}{*}{Lifetime {[}s{]}}}                                 & \multicolumn{1}{c|}{$\mu_{T}$}         & \multicolumn{1}{c|}{2.751}   & \multicolumn{1}{c|}{2.925}   & 2.795   \\
\multicolumn{1}{c|}{}                         & \multicolumn{2}{c|}{}                                                                  & \multicolumn{1}{c|}{$\sigma_{T}$}         & \multicolumn{1}{c|}{0.632}   & \multicolumn{1}{c|}{0.708}   & 0.631   \\ \cline{2-7} 
\multicolumn{1}{c|}{}                         & \multicolumn{2}{c|}{New number}                                                        & \multicolumn{4}{c}{see as Table $\textrm{II}$}                                                                                  \\ \cline{2-7} 
\multicolumn{1}{c|}{}                         & \multicolumn{1}{c|}{\multirow{10}{*}{Initial power [dB]}} & \multirow{5}{*}{{[}0-50{]} ns}   & \multicolumn{1}{c|}{$p$}         & \multicolumn{1}{c|}{0.184}   & \multicolumn{1}{c|}{-0.197}  & -0.097  \\
\multicolumn{1}{c|}{}                         & \multicolumn{1}{c|}{}                                &                                 & \multicolumn{1}{c|}{$q$}         & \multicolumn{1}{c|}{-61.191} & \multicolumn{1}{c|}{-45.342} & -48.142 \\
\multicolumn{1}{c|}{}                         & \multicolumn{1}{c|}{}                                &                                 & \multicolumn{1}{c|}{$\xi_r$}         & \multicolumn{1}{c|}{-0.112}  & \multicolumn{1}{c|}{0.025}   & 0.067   \\
\multicolumn{1}{c|}{}                         & \multicolumn{1}{c|}{}                                &                                 & \multicolumn{1}{c|}{$\sigma_r$}         & \multicolumn{1}{c|}{4.089}   & \multicolumn{1}{c|}{4.667}   & 4.103   \\
\multicolumn{1}{c|}{}                         & \multicolumn{1}{c|}{}                                &                                 & \multicolumn{1}{c|}{$\mu_r$}         & \multicolumn{1}{c|}{-2.908}  & \multicolumn{1}{c|}{-8.231}  & -7.421  \\ \cline{3-7} 
\multicolumn{1}{c|}{}                         & \multicolumn{1}{c|}{}                                & \multirow{5}{*}{{[}51-end{]} ns} & \multicolumn{1}{c|}{$p$}         & \multicolumn{1}{c|}{-0.015}  & \multicolumn{1}{c|}{-0.002}  & -0.002  \\
\multicolumn{1}{c|}{}                         & \multicolumn{1}{c|}{}                                &                                 & \multicolumn{1}{c|}{$q$}         & \multicolumn{1}{c|}{-55.071} & \multicolumn{1}{c|}{-53.371} & -54.073 \\
\multicolumn{1}{c|}{}                         & \multicolumn{1}{c|}{}                                &                                 & \multicolumn{1}{c|}{$\xi_r$}         & \multicolumn{1}{c|}{0.135}   & \multicolumn{1}{c|}{0.121}   & 0.105   \\
\multicolumn{1}{c|}{}                         & \multicolumn{1}{c|}{}                                &                                 & \multicolumn{1}{c|}{$\sigma_r$}         & \multicolumn{1}{c|}{3.299}   & \multicolumn{1}{c|}{3.848}   & 3.772   \\
\multicolumn{1}{c|}{}                         & \multicolumn{1}{c|}{}                                &                                 & \multicolumn{1}{c|}{$\mu_r$}         & \multicolumn{1}{c|}{-3.294}  & \multicolumn{1}{c|}{-4.021}  & -3.463  \\ \cline{2-7} 
\multicolumn{1}{c|}{}                         & \multicolumn{2}{c|}{\multirow{2}{*}{Initial delay {[}ns{]}}}                           & \multicolumn{1}{c|}{$\alpha_d$}         & \multicolumn{1}{c|}{1.311}   & \multicolumn{1}{c|}{1.141}   & 1.028   \\
\multicolumn{1}{c|}{}                         & \multicolumn{2}{c|}{}                                                                  & \multicolumn{1}{c|}{$\beta_d$}         & \multicolumn{1}{c|}{81.621}  & \multicolumn{1}{c|}{62.431}  & 85.083  \\ \cline{2-7} 
\multicolumn{1}{c|}{}                         & \multicolumn{2}{c|}{\multirow{2}{*}{Power evolution [dB]}}                                  & \multicolumn{1}{c|}{$\mu_p$}         & \multicolumn{1}{c|}{0.001}   & \multicolumn{1}{c|}{-0.034}  & -0.032  \\
\multicolumn{1}{c|}{}                         & \multicolumn{2}{c|}{}                                                                  & \multicolumn{1}{c|}{$\sigma_p$}         & \multicolumn{1}{c|}{0.747}   & \multicolumn{1}{c|}{0.864}   & 0.947   \\ \cline{2-7} 
\multicolumn{1}{c|}{}                         & \multicolumn{2}{c|}{\multirow{3}{*}{Delay evolution [ns]}}                                  & \multicolumn{1}{c|}{$k_d$}         & \multicolumn{1}{c|}{0.039}   & \multicolumn{1}{c|}{0.028}   & 0.025   \\
\multicolumn{1}{c|}{}                         & \multicolumn{2}{c|}{}                                                                  & \multicolumn{1}{c|}{$\mu_d$}         & \multicolumn{1}{c|}{1.625}   & \multicolumn{1}{c|}{-0.354}  & 0.132   \\
\multicolumn{1}{c|}{}                         & \multicolumn{2}{c|}{}                                                                  & \multicolumn{1}{c|}{$\sigma_d$}         & \multicolumn{1}{c|}{6.361}   & \multicolumn{1}{c|}{4.203}   & 4.403   \\ \cline{2-7} 
\multicolumn{1}{c|}{}                         & \multicolumn{2}{c|}{Number within cluster}                                             & \multicolumn{4}{c}{see as Table $\textrm{III}$}                                                                                  \\ \hline
\multicolumn{1}{c|}{\multirow{6}{*}{C-MPCs}}  & \multicolumn{1}{c|}{\multirow{4}{*}{Power decay [dB]}}  & \multirow{2}{*}{{[}0-50{]} ns}   & \multicolumn{1}{c|}{$a$}         & \multicolumn{1}{c|}{0.009}   & \multicolumn{1}{c|}{-0.038}  & -0.074  \\
\multicolumn{1}{c|}{}                         & \multicolumn{1}{c|}{}                                &                                 & \multicolumn{1}{c|}{$b$}         & \multicolumn{1}{c|}{-64.72}  & \multicolumn{1}{c|}{-60.27}  & -59.62  \\ \cline{3-7} 
\multicolumn{1}{c|}{}                         & \multicolumn{1}{c|}{}                                & \multirow{2}{*}{{[}51-end{]} ns} & \multicolumn{1}{c|}{$a$}         & \multicolumn{1}{c|}{-0.003}  & \multicolumn{1}{c|}{-0.004}  & -0.004  \\
\multicolumn{1}{c|}{}                         & \multicolumn{1}{c|}{}                                &                                 & \multicolumn{1}{c|}{$b$}         & \multicolumn{1}{c|}{-64.28}  & \multicolumn{1}{c|}{-62.01}  & -62.9   \\ \cline{2-7} 
\multicolumn{1}{c|}{}                         & \multicolumn{2}{c|}{\multirow{2}{*}{Fading [dB]}}                                           & \multicolumn{1}{c|}{$\mu_c$}         & \multicolumn{1}{c|}{-2.764}  & \multicolumn{1}{c|}{-3.88}   & -3.28   \\
\multicolumn{1}{c|}{}                         & \multicolumn{2}{c|}{}                                                                  & \multicolumn{1}{c|}{$\sigma_c$}         & \multicolumn{1}{c|}{5.654}   & \multicolumn{1}{c|}{6.035}   & 5.846   \\ \hline
\end{tabular}}
\end{table}

In this subsection, we summarize statistic parameters mentioned before as Table $\textrm{IV}$, and summarize model implementation steps as Fig.12. First, input the simulation time \( T \) and snapshot sampling rate \( r \) to determine the total number of simulation snapshots \( K \), and specify the sensing directions. Different sensing directions adopt different channel statistical parameters. Then, initialize the number of S-MPCs in the channel and assign each S-MPC with its lifetime, initial power, and initial delay based on fitted statistical distributions. At the next time step, discard dead S-MPCs based on their lifetime, while counting the remaining number of S-MPCs \( N_r \) and the number of multipaths \( N_s \) within cluster. Note that if the remaining S-MPCs are zero, indicating that all old S-MPCs have disappeared, reinitialize them; otherwise, evolve the remaining S-MPCs over time, including updating their power and delay according to the fitted functions, then generate new S-MPCs based on the distribution model, including their number, initial power, and initial delay, and perform clustering. Iterate through the above steps repeatedly until the simulation time is met. Through iteratively repeating the model generation steps, we can obtain dynamically evolving vehicular ISAC channels that extend infinitely along time.

\subsection{Validation}

With the proposed model, a dynamic channel impulse responses (CIR) over successive time can be generated for vehicular ISAC channels. We generate CIRs for the duration of 1200 s in three directions. Considering the complexity of the model, the average running time for generating CIR of 1s is approximately 0.026 s (in MATLAB 2022, with 64 GB RAM and Core-i5 CPU), which shows that the model has very low complexity of simulation. To further validate the model, second-order statistics, i.e., the root-mean-square delay spread (RMSDS) is employed. Fig. 13 shows the CDF comparisons of RMSDS between measured and simulated ISAC channels in different directions. It can be observed that the simulation is fairly close to the measurements for both front, left and right directions. Note that the simulated CDFs tend to be slightly wider than the measured ones. A possible explanation is that the specific locations of scatterers are not taken into account in simulation, resulting in a more generalized RMSDS range. Overall, the simulated ISAC channel based on the proposed model can reflect the dynamic characteristics of actual channel well, and the research on dynamic ISAC channel models can be the foundation for vehicular ISAC system design and performance evaluation.

\begin{figure*}[tbp]
\centering
\subfigure[]{\includegraphics[width=2.3in]{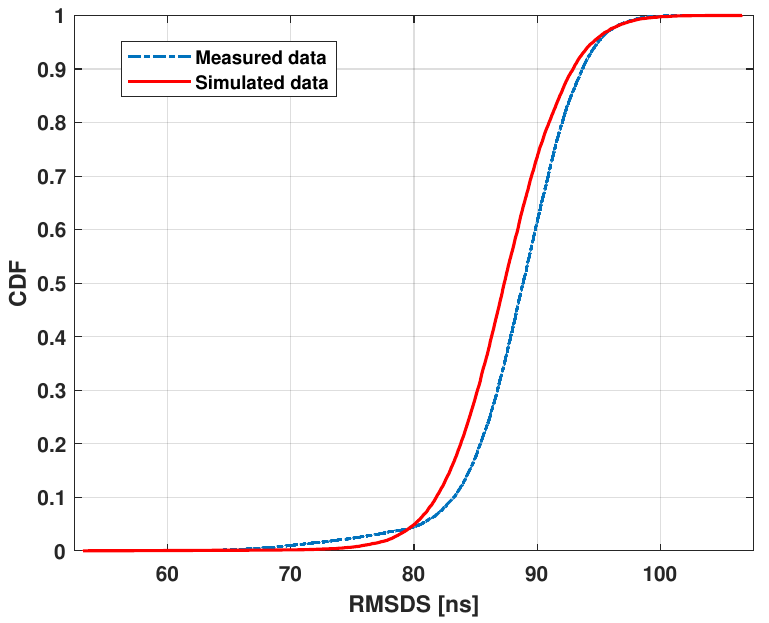}}
\subfigure[]{\includegraphics[width=2.3in]{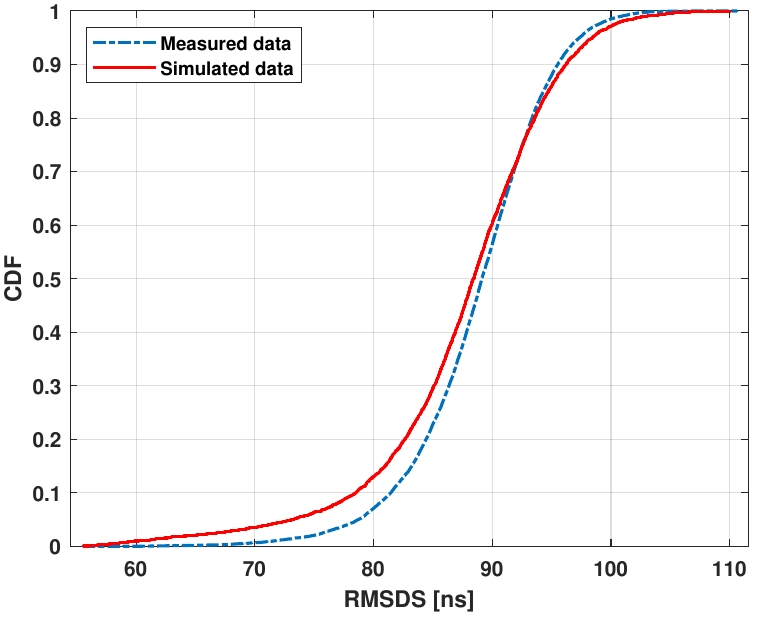}}
\subfigure[]{\includegraphics[width=2.3in]{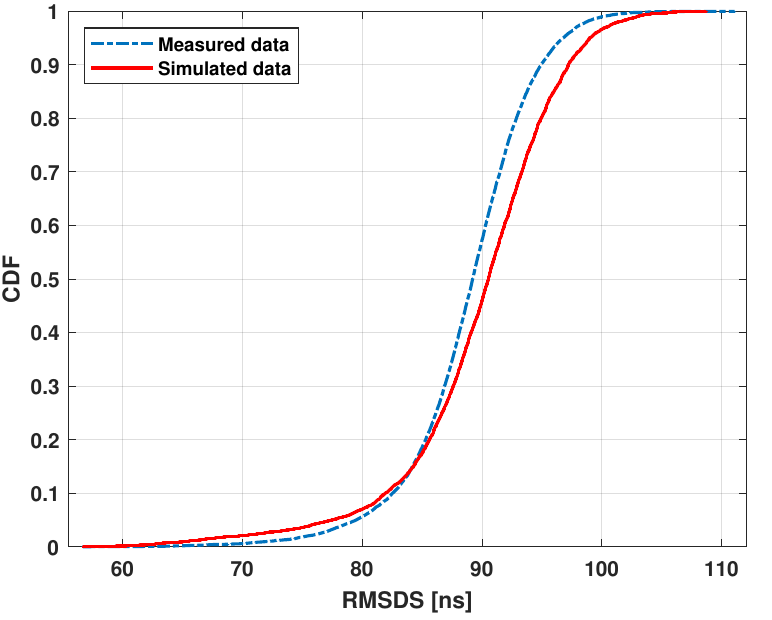}}
\caption{RMSDS between measured and simulated ISAC channels. (a) Front. (b) Left. (c) Right.}
\label{fig}
\end{figure*}

\section{Conclusion}

In this paper, a dynamic statistical channel model is proposed for vehicular ISAC scenarios, which is composed of S-MPCs and C-MPCs. With actual vehicular ISAC channel measurements at 28 GHz, this paper uses a MCD-based tracking algorithm to identify S-MPCs and C-MPCs, and analyzes time-varying sensing characteristics of front, left, and right directions in complicated traffic scenarios. To model the dynamic evolution process, number of new S-MPCs, lifetimes, initial power and delay positions, dynamic variations within their lifetimes, clustering, power decay and fading of C-MPCs are statistically characterized, which are fitted to the best theoretical distributions. In addition, this paper shows the implementation of dynamic model, and validates the model by comparing key statistics of simulations with measurements. The proposed model can generate dynamic behaviors of MPCs in vehicular ISAC scenarios and can be used for vehicular ISAC system design and performance evaluation.


\bibliographystyle{IEEEtran}

\bibliography{2021_TVT}





\end{document}